\documentclass[%
 aip,
 amsmath,amssymb,
 reprint,
 normalem,%
]{revtex4-1}

\usepackage{graphicx}
\usepackage{dcolumn}
\usepackage{bm}
\usepackage{dsfont}
\usepackage{xcolor}
\usepackage{ulem}
\usepackage[utf8]{inputenc}
\usepackage[T1]{fontenc}
\usepackage{mathptmx}
\usepackage{mathtools}
\usepackage{etoolbox}
\usepackage{tabularx}
\usepackage{placeins}
\usepackage{xspace}
\usepackage{booktabs}
\usepackage{multirow}
\usepackage[hidelinks]{hyperref}

\def\tddftb{LC-TD-DFTB\xspace}
\def\dftb{LC-DFTB\xspace}
\def\fmo{FMO-LC-TDDFTB\xspace}

\makeatletter
\def\@email#1#2{%
 \endgroup
 \patchcmd{\titleblock@produce}
  {\frontmatter@RRAPformat}
  {\frontmatter@RRAPformat{\produce@RRAP{*#1\href{mailto:#2}{#2}}}\frontmatter@RRAPformat}
  {}{}
}%
\makeatother
\begin{document}

\title[]{Long-range Corrected Fragment Molecular Orbital Density-Functional Tight-binding Method for Excited States in Large Molecular Systems}

\author{Richard Einsele}
\author{Joscha Hoche}
\author{Roland Mitri\'c}
\email{roland.mitric@uni-wuerzburg.de}
\affiliation{Institut für Physikalische und Theoretische Chemie, Julius-Maximilians-Universität Würzburg, Emil-Fischer-Strasse 42, 97074 Würzburg, Germany}
\date{\today}% It is always \today, today,

\begin{abstract}
Herein, we present a new method to efficiently calculate electronically excited states in large molecular assemblies, consisting of hundreds of molecules. For this purpose, we combine the long-range corrected tight-binding density-functional fragment molecular orbital method (FMO-LC-DFTB) with an excitonic Hamiltonian, which is constructed in the basis of locally excited and charge-transfer configuration state functions calculated for embedded monomers and dimers and accounts explicitly for the electronic coupling between all types of excitons. We first evaluate both the accuracy and efficiency of our fragmentation approach for molecular dimers and aggregates by comparing with the full \tddftb method. The comparison of the calculated spectra of an anthracene cluster shows a very good agreement between our method and the \tddftb reference. The effective computational scaling of our method has been explored for anthracene clusters and for perylene bisimide aggregates. We demonstrate the applicability of our method by the calculation of the excited state properties of pentacene crystal models consisting of up to 319 molecules. Furthermore, the participation ratio of the monomer fragments to the excited states is analyzed by the calculation of natural transition orbital (NTO) participation numbers, which are verified by the hole and particle density for a chosen pentacene cluster. The use of our \fmo method will allow for future studies of excitonic dynamics and charge transport to be performed on complex molecular systems consisting of thousands of atoms.
\end{abstract}

\maketitle

\section{Introduction}
In materials science, solid-state physics, biochemistry and many other branches of molecular sciences, the study of excited states is essential for understanding photochemical and photophysical processes as well as the function of molecular materials. Simulations of the excited state properties necessitate an accurate description of the electronic structure of molecular systems, which was first accomplished with ab-initio quantum chemical methods\cite{friesner_new_1991,gonzalez_progress_2012}.

However, investigating excited states of extended systems in the framework of the ab-initio approach requires very long calculation times due to the high computational scaling with the system size. With the development of the time-dependent density functional theory (TD-DFT), a reasonably fast method of calculating the excited state properties of relatively large molecular systems (over 100 atoms) became available \cite{casida_time-dependent_1995,casida_progress_2012,laurent_td-dft_2013}. 
New developments of approximative methods for the prediction of the nature of the excited states, including linear-scaling TD-DFT\cite{wu_linear-scaling_2011}, subsystem density functional theory\cite{jacob_subsystem_2014} and multilayer QM/MM approaches (ONIOM)\cite{dapprich_new_1999,vreven_combining_2006}, have further increased the applicability of TD-DFT to large polyatomic systems. However, despite the multitude of recent developments, the calculation of the excited states of molecular systems consisting of thousand atoms or more is still not possible even with TD-DFT.

The semiempirical quantum mechanical (SQM) methods, which approximate the wavefunction or density functional based methods, provide a further alternative for investigating the excited states of molecular systems. In the various SQM theories, a minimal basis set is used and the electronic integrals are approximated by the partial or complete neglect of the differential overlap between the atomic basis functions, which leads to a sizable increase in the computational efficiency  \cite{thiel_semiempirical_2014,christensen_semiempirical_2016}. Whereas these approaches have been studied extensively in the 1970s--1990s\cite{dewar_ground_1977,dewar_development_1985,stewart_optimization_1989,kolb_beyond_1993,thiel_extension_1996,weber_orthogonalization_2000}, the development of the self-consistent charge (SCC) density functional tight-binding (DFTB) method\cite{porezag_construction_1995,seifert_calculations_1996,elstner_self-consistent-charge_1998} has renewed the interest in semiempirical approaches in the last 20 years. Niehaus \textit{et al.}\cite{niehaus_tight-binding_2001} adapted the time-dependent density functional theory to the tight-binding formalism, enabling the calculation of excited state properties of large molecular assemblies of hundreds of atoms. The introduction of long-range corrections to the DFTB framework facilitated the investigation of excited state properties of molecular systems involving charge-transfer states \cite{niehaus_importance_2005,humeniuk_long-range_2015,lutsker_implementation_2015}. The DFTB and TD-DFTB methods have been implemented in various software packages including DFTB+\cite{hourahine_dftb_2020}, DFTBaby\cite{humeniuk_dftbaby_2017}, ADF\cite{te_velde_chemistry_2001}, CP2K\cite{hutter_cp2k_2014}, and hotbit\cite{koskinen_density-functional_2009}, which can be employed to predict the excited state properties of polyatomic systems of over 1000 atoms. 

Recently, Grimme developed the simplified Tamm-Dancoff approximation\cite{grimme_simplified_2013} (sTDA) for the calculation of excited-state spectra of large molecules, which is based on a ground state DFT calculation, followed by the semiempirical treatment of the integrals involved in the linear-response TD-DFT calculation, similar to the TD-DFTB approach. Furthermore, this method was extended to the tight-binding formalism (sTDA-xTB)\cite{grimme_ultra-fast_2016} by the parametrization of the ground-state Hamiltonian to reduce the computational requirements of the previous models. Further developments of this approach include the GFN-xTB methods by Grimme and coworkers, which can be employed to predict the ground state properties and structures of a multitude of molecular systems \cite{grimme_robust_2017,bannwarth_gfn2-xtbaccurate_2019,bannwarth_extended_2021}. Additional notable semiempirical methods, which can be used to predict the excited state properties of large molecular systems are MNDO\cite{dewar_ground_1977}, AM1\cite{dewar_development_1985}, PMx\cite{stewart_optimization_1989} and OMx\cite{weber_orthogonalization_2000}. 

The computational efficiency and applicability of all above methods can be further enhanced by using fragmentation-based approaches allowing the simulation of large molecular systems of over a million atoms. Prominent examples include the fragment molecular orbital\cite{kitaura_fragment_1999,nakano_fragment_2000,nakano_fragment_2002,fedorov_multilayer_2005,mochizuki_configuration_2005,tsuneyuki_molecular_2009,brorsen_fully_2011,fujino_fragment_2015,tanaka_electron-correlated_2014} (FMO) method and the divide-and-conquer\cite{yoshikawa_novel_2013,nishizawa_three_2016,nakai_development_2017,yoshikawa_gpuaccelerated_2019,nishimura_quantum_2021} (DC) scheme. In these fragmentation approaches, a partitioning scheme divides the system into multiple fragments. The properties of the complete system are then obtained by combining the results of the different fragments. In the case of the FMO method, the energy of the fragments -- the monomers and fragment pairs -- is calculated iteratively while considering the electrostatic embedding potential (ESP) resulting from the interaction of all fragments. The number of fragment pairs is determined by the closest atomic interfragment distance. The FMO and the DC approach have previously been employed to investigate complex molecular systems, such as proteins\cite{sawada_role_2010}, polymers\cite{nakata_derivatives_2014} or optoelectronic materials\cite{uratani_simulating_2020}. Furthermore, Nishimoto, Fedorov and Irle combined the FMO method with the density functional tight-binding approach (FMO-DFTB) to enable the study of even larger systems \cite{nishimoto_density-functional_2014,nishimoto_fmo-dftb_2021}. The FMO-DFTB method has been extended to include analytical ground-state gradients\cite{nishimoto_large-scale_2015}, the polarizable continuum model\cite{nishimoto_fragment_2016}, periodic boundary conditions\cite{nishimoto_fragment_2021}, the long-range correction\cite{vuong_fragment_2019} and many other theoretical concepts\cite{nishimoto_third-order_2015,nishimoto_three-body_2017,nishimoto_adaptive_2018,fedorov_partition_2020}. It has been succesfully employed to study the properties of large molecular systems, such as the charge transport in covalent organic frameworks \cite{kitoh-nishioka_multiscale_2017,kitoh-nishioka_linear_2021}.

While these fragmentation-based methods provide excellent tools to calculate the ground state properties of molecular aggregates, the study of excited states is limited to only local excitations within the fragments\cite{komoto_development_2019,komoto_large-scale_2020,uratani_fast_2020,uratani_non-adiabatic_2020,uratani_trajectory_2021}. Recently, several approaches have been developed to investigate the excited states properties of molecular clusters using an excitonic Hamiltonian, consisting of local excitations (LEs) and charge-transfer states (CTs)\cite{green_excitonic_2021,li_ab_2017,wen_fragmentation-based_2017}. Fujita and Mochizuki combined the multilayer FMO\cite{fedorov_multilayer_2005} and the transition density fragment interaction\cite{fujimoto_transition-density-fragment_2012,fujimoto_theoretical_2013} method and introduced an excitonic Hamiltonian to calculate nonlocal excitations in molecular clusters. A basis consisting of fragment configuration state functions (CSFs), which define the locally excited and charge-transfer states, is used to construct the Hamiltonian\cite{fujita_development_2018}. 

In this work, we combine the FMO-LC-DFTB method with the construction of an excitonic Hamiltonian to calculate the excited states of large molecular systems consisting of hundreds of molecules. As a basis for the description of the excited states of large molecular assemblies, we use locally excited and charge-transfer states, which are calculated for monomers and pair dimers employing the \tddftb approach. While the diagonal elements of the Hamiltonian are represented by the energies of the basis states, the couplings between the basis states, which represent the off-diagonal matrix elements of the Hamiltonian, are calculated by utilizing the tight-binding formalism for the two-electron four-center integrals. The excited state energies of the full system are given by the diagonalization of the excitonic Hamiltonian. 
The accuracy and the efficiency of the proposed theory has been tested by the comparison with the full \tddftb method for molecular dimers and large assemblies of molecules. The effective computational scaling was determined by the calculation of large clusters of anthracene and perylene bisimide aggregates. In addition, the applicability of our approach was confirmed by the calculation of the excited state properties of pentacene crystal models. The \fmo method was implemented in our own software package \mbox{DIALECT}\cite{noauthor_dialect_2022}, which is publicly available on Github and was written in the Rust programming language. 

Although Rust is still a very young language --- with the first stable version (1.0) released in 2015 ---  its potential for scientific computing applications becomes increasingly apparent\cite{perkel_why_2020}. The Rust programming language does not only have a strong focus on the code and memory safety, but also shows performance comparable to the C, C++ and Fortran languages. Additionally, Rust offers in our opinion a friendlier syntax and provides a strongly growing repository of development tools. Therefore, we expect Rust to gain high popularity also in the field of quantum chemistry in the coming years.

The present work is structured as follows: In Sec. \ref{sec:methodology}, the methodological framework of our approach is outlined. Subsequently, in Sec. \ref{sec:computational_procedure}, the computational procedure and the different steps of a \fmo calculation are illustrated. The application of the proposed theory to pentacene and the results of the benchmark calculations regarding the accuracy and the effective scaling of the \fmo method are presented in Sec. \ref{sec:results}. Finally, the conclusions and outlook are given in Sec. \ref{sec:conclusions}.
\newcommand\numberthis{\addtocounter{equation}{1}\tag{\theequation}}
\section{Methodology}\label{sec:methodology}

In the following, we will present a general formulation of the fragment molecular orbital (FMO) LC-TD-DFTB. As a starting point, we review the theory behind LC-TD-DFTB\cite{humeniuk_long-range_2015, lutsker_implementation_2015}. This is followed by the formulation of the FMO-ansatz for the description of the electronic ground state. Subsequently, we present how the molecular orbitals of the total system can be calculated and introduce a theoretical approach for calculating electronically excited states. This is concluded with a description of the necessary Hamiltonian matrix elements and the computational procedure that is used in our implementation. 
Throughout the paper we use atomic units and the following notation convention: atoms are denoted as uppercase letters from A-H and molecular fragments from I-Z without indices. Uppercase letters with indices denote matrix elements and bold uppercase letters matrices. For molecular orbital (atomic orbital) indices lowercase (greek) letters are used. 
It should be noted that in the following we use the terms fragment and monomer synonymously to refer to a molecular unit. Likewise, we use (molecular) cluster and aggregate synonymously to refer to a large number of molecular units that are not covalently linked. 

\subsection{Long-range corrected DFTB (LC-DFTB)}
As shown previously, the formalism of LC-DFTB can be derived by a Taylor expansion of a long-range corrected full density functional (e.g. LC-PBE) around a reference density that is given by a sum of atomic densities, as described in detail elsewhere \cite{humeniuk_long-range_2015, lutsker_implementation_2015}. In the following, we will only provide a brief summary of the working equations. After employing the usual tight-binding (TB) approximations\cite{elstner_self-consistent-charge_1998} the total energy in SCC LC-DFTB is given by:
\begin{align}
E=& \sum_{\mu \nu} P_{\mu \nu} H_{\mu \nu}^{0}  \tag{band structure energy}\\
&+ \frac{1}{2} \sum_{\mu, \sigma, \lambda, \nu} \Delta P_{\mu \sigma} \Delta P_{\lambda \nu}(\mu \sigma \mid \lambda \nu) \tag{Coulomb energy} \\
&-\frac{1}{4} \sum_{\mu, \sigma, \lambda, \nu} \Delta P_{\mu \sigma}  \Delta P_{\lambda \nu}(\mu \lambda \mid \sigma \nu)_{\operatorname{lr}}\tag{exchange energy}\\
& + \sum_{A, B} V_{A B}^{\mathrm{rep}}\left(R_{A B}\right)\tag{repulsion energy} 
\label{eq:dftb_energy}
\end{align}
Here $P_{\mu \nu}$ denotes the electron density matrix elements and $H_{\mu \nu}^{0}$ the one-electron Hamiltonian matrix elements. The electron density difference matrix is defined as $\Delta \mathbf{P} = \mathbf{P} - \mathbf{P}^0$, where $\mathbf{P}^0$ is the diagonal reference electron density matrix. The deviation of the Mulliken charge $q_A$ on Atom A from the charge of the neutral Atom $q^0_A$ is given by
\begin{align}
    \Delta q_A &= q_A - q^0_A \\
    &= \sum_{\mu \in A} \sum_{\nu} \left[\mathbf{P}_{\mu\nu} S_{\mu\nu} - \mathbf{P}^0_{\mu\nu} S_{\mu\nu} \right],
\end{align}
where $S_{\mu\nu}$ represents a matrix element of the overlap matrix.

Applying the tight-binding approximations to the two-electron integrals in the Coulomb and exchange part of the energy and neglecting all 3- and 4-center integrals gives the following expressions: 
\begin{align*}
(\mu \lambda \mid \sigma \nu) &=\iint \phi_{\mu}(r_1) \phi_{\lambda}(r_1) \frac{1}{r_{12}} \phi_{\sigma}(r_2) \phi_{\nu}(r_2) \mathrm{d} 1 \mathrm{d} 2 \\
& \approx \sum_{A, B} \gamma_{A B} q_{A}^{\mu \lambda} q_{B}^{\sigma \nu}  \numberthis{}{}
\label{eq:2e_approx_ao}
\end{align*}
with the transition charges on atom $A$ (in the atomic orbital basis) defined as
\begin{equation}
q_{A}^{\mu \lambda}=\frac{1}{2}(\delta(\mu \in A)+\delta(\lambda \in A)) S_{\mu \nu}
\label{eq:q_between_aos}
\end{equation}
The $\bm{\gamma}$-matrices for charge fluctuation interactions can be calculated assuming that the charge fluctuations can be represented by spherical Gaussian functions, leading to
\begin{equation}
\gamma_{A B}=\frac{\operatorname{erf}\left(C_{A B} R\right)}{R},
\label{eq:gamma}
\end{equation}
where $R$ is the distance between the atomic centers $A$ and $B$ and $C_{A B}=(2\left(\sigma_A^2+\sigma_B^2\right))^{-\frac{1}{2}}$ depends on the widths $\sigma_A$ and $\sigma_B$ of the charge clouds on the two atoms. The widths are determined by the atom-specific Hubbard parameters $U_A$ as $\sigma_A= (\sqrt{\pi} U_A)^{\frac{1}{2}}$.
In LC-DFTB the Coulomb potential is separated into a long-range and a short range part where the position of the smooth transition between the two regimes is controlled by the long-range radius $R_{\mathrm{lr}}$
\begin{equation}
\frac{1}{r}=\underbrace{\frac{1-\operatorname{erf}\left(\frac{r}{R_{\mathrm{lr}}}\right)}{r}}_{\text {short range }}+\underbrace{\frac{\operatorname{erf}\left(\frac{r}{R_{\mathrm{lr}}}\right)}{r}}_{\text {long range }}.
\end{equation}
The short range term is already included by the local exchange-correlation functional employed in LC-DFTB. The electron integrals of the screened Coulomb potential for the long-range contribution can be calculated as 
\begin{align*}
(\mu \lambda \mid \sigma \nu)_{\mathrm{lr}} &=\iint \phi_{\mu}(r_1) \phi_{\lambda}(r_1) \frac{\operatorname{erf}\left(\frac{r_{12}}{R_{\mathrm{lr}}}\right)}{r_{12}} \phi_{\sigma}(r_2) \phi_{\nu}(r_2) \mathrm{d} 1 \mathrm{d} 2 \\
& \approx \sum_{A, B} \gamma_{A B}^{\operatorname{lr}} q_{A}^{\mu \lambda} q_{B}^{\sigma \nu}. \numberthis{}{}
\label{eq:2e_approx_ao_lr}
\end{align*}
 The long-range $\gamma$-matrix is defined similarly to the $\gamma$-matrix in eq. \ref{eq:gamma},
\begin{equation}
\gamma_{A B}^{\operatorname{lr}}=\frac{\operatorname{erf}\left(C_{A B}^{\mathrm{lr}} R\right)}{R},
\end{equation}
where
\begin{equation}
C_{A B}^{\mathrm{lr}}=\frac{1}{\sqrt{2\left(\sigma_A^2+\sigma_B^2+\frac{1}{2} R_{\mathrm{lr}}^2\right)}}
\end{equation}
depends on the range-separation parameter $R_{\mathrm{lr}}$.
The transformation of the two-electron repulsion integrals into molecular orbital (MO) basis leads to the following expressions:
\begin{equation}
    (pq \mid rs ) \approx \sum_{A} \sum_{B} q_{A}^{pq} \gamma_{AB} q_{B}^{rs}
    \label{eq:2e_approx_mo}
\end{equation}
and 
\begin{equation}
    (pq \mid rs )_{\mathrm{lr}} \approx \sum_{A} \sum_{B} q_{A}^{pq} \gamma_{AB}^{\mathrm{lr}} q_{B}^{rs}
    \label{eq:2e_approx_mo_lr}
\end{equation}
The atom-centered transition charges between MOs are defined as: 
\begin{equation}
q_{A}^{i j}=\frac{1}{2} \sum_{\mu \in A} \sum_{\nu}\left(c_{\mu}^{i} c_{\nu}^{j}+c_{\nu}^{i} c_{\mu}^{j}\right) S_{\mu \nu}
\label{eq:q_between_mos}
\end{equation}
Variational minimization of the LC-DFTB energy with respect to the molecular orbitals leads to the corresponding Hamiltonian, that is defined as: 
\begin{align*}
H_{\mu \nu}^{\mathrm{SCC}} &=H_{\mu \nu}^{0}+\frac{1}{2} S_{\mu \nu} \sum_{C}\left(\gamma_{A C}+\gamma_{B C}\right) \Delta q_{C} \\
&-\frac{1}{8} \sum \Delta P_{\alpha \beta} S_{\mu \alpha} S_{\beta \nu}\left(\gamma_{\mu \beta}^{\mathrm{lr}}+\gamma_{\mu \nu}^{\mathrm{lr}}+\gamma_{\alpha \beta}^{\mathrm{lr}}+\gamma_{\alpha \nu}^{\mathrm{lr}}\right) \numberthis{}{} \label{eq:scc_hamiltonian}
\end{align*}
The ground state MOs and their energy are obtained by a self consistent procedure for solving the following general eigenvalue problem
\begin{equation}
    \mathbf{Hc} = \epsilon\mathbf{Sc}
\end{equation}

Being a semiempirical theory, (LC)-DFTB is heavily dependent on a parametrization for the electronic Hamiltonian, $H_{\mu \nu}^{0}$, the overlap matrix $S_{\mu \nu}$ and the repulsive potentials  $V_{A B}^{\mathrm{rep}}$. The parametrization usually starts with the computation of pseudoorbitals, the tabulation of Slater-Koster files and ends with the fitting of repulsive potentials \cite{elstner_density_2014,koskinen_density-functional_2009,spiegelman_density-functional_2020}. A benchmark of the used LC-DFTB parametrization for organic and biological molecules is given in Ref.\cite{vuong_parametrization_2018}.

\subsection{Fragment molecular orbital LC-DFTB (FMO-LC-DFTB)}
The fragment molecular orbital method in combination with DFTB was developed by Nishimoto, Fedorov and Irle\cite{nishimoto_density-functional_2014}. Recently, it was adopted to LC-DFTB by Niehaus and Irle\cite{vuong_fragment_2019} and this theory forms the basis for our ground-state calculations. Therefore, we will only restate the most important equations in the following and refer to Ref. \cite{nishimoto_density-functional_2014} and Ref. \cite{vuong_fragment_2019} for a more detailed description. Only the case of molecular clusters --- molecules that are not covalently connected --- is considered in this work, and therefore, we do not introduce the hybrid orbital projection operator \cite{nakano_fragment_2000}.
The total energy of the entire system in the self-consistent FMO-LC-DFTB is given by
\begin{equation}
E=\sum_{I}^{N} E_{I}+ \frac{1}{2} \sum_{I}^{N} \sum_{J}^N \left(E_{I J}-E_{I}-E_{J}\right).
\label{eq:fmo_energy}
\end{equation}
Here $E_I$ ($E_J$) are the energies of fragment $I$ ($J$) and $E_{IJ}$ is the energy of the pair $IJ$. The total number of fragments is denoted as $N$, while we will use $N_I$ for the number of atoms in fragment $I$. 
The energies $E_X$ where $X$ is either $I$ or $IJ$ can be further separated into a part that depends only on the isolated molecule and another one that accounts for the environment of the molecule:
\begin{equation}
E_{X}=E_{X}^{\prime}+E_{X}^{\mathrm{em}}
\label{eq:frag_energy}
\end{equation}
The former is the so-called internal energy, $E_{X}^{\prime}$, which is equivalent to the total energy calculated in conventional LC-DFTB of entity $X$. The embedding energy, $E_{X}^{\mathrm{em}}$, accounts for the Coulomb interaction between the entity $X$ and all other fragments: 
\begin{equation}
E_{X}^{\mathrm{em}}=\sum_{A \in X} \sum_{K \neq X}^{N} \sum_{C \in K} \gamma_{A C} \Delta q_{A}^{X} \Delta q_{C}^{K}
\label{eq:embedding_energy_frag}
\end{equation}
Inserting eq. \ref{eq:frag_energy} and eq. \ref{eq:embedding_energy} in eq. \ref{eq:fmo_energy} allows to rewrite the total FMO energy as: 
\begin{equation}\label{eq:total_fmo_energy}
E=\sum_{I}^{N} E_{I}^{\prime}+ \frac{1}{2} \sum_{I}^{N} \sum_{J}^N \left(E_{I J}^{\prime}-E_{I}^{\prime}-E_{J}^{\prime}\right)+ \frac{1}{2} \sum_{I}^{N} \sum_{J}^{N} \Delta E_{I J}^{\mathrm{em}}
\end{equation}
The last term on the r.h.s. is the difference in embedding energy of the pair and corresponding embedded but noninteracting monomers. 
\begin{equation}\label{eq:embedding_energy}
\Delta E_{I J}^{\mathrm{em}} = E_{I J}^{\mathrm{em}} - E_{I}^{\mathrm{em}} - E_{J}^{\mathrm{em}} = \sum_{A \in I J} \sum_{K \neq I, J}^{N} \sum_{C \in K} \gamma_{A C} \Delta \Delta q_{A}^{I J} \Delta q_{C}^{K}
\end{equation}
Here $\Delta \Delta q_{A}^{I J}$ denotes the difference between charges of pair $IJ$ and charges of fragments $I$ and $J$ for atom $A$:
\begin{align}
\Delta \Delta q_{A}^{I J}   &= \Delta q_{A}^{I J} - ( \Delta q_{A}^{I} \oplus \Delta q_{A}^{J}) \\
\Delta q_{A}^{I} \oplus \Delta q_{A}^{J}  &= \Delta q_{A}^{I}  \text { for } A \in I \text { and } A \notin J,\\
                            &= \Delta q_{A}^{J} \text { for } A \in J \text { and } A \notin I, 
\end{align}
We employ the electrostatic dimer (ES-DIM) approximation \cite{nakano_fragment_2002} for pairs, in which the fragments are so far separated, that their orbital overlap will be zero. The energy of the far-separated pairs is given by 
\begin{equation}\label{eq:esdim_energy}
E_{I J}^{\prime} = E_I^{\prime}+E_J^{\prime}+\sum_{A \in I} \sum_{B \in J} \gamma_{A B} \Delta q_A^I \Delta q_B^J.
\end{equation}
It should be noted that although it is possible to separate the total energy into internal and embedding energies, the tight-binding Hamiltonian of an entity (\textit{cf.} eq. \ref{eq:scc_hamiltonian}) also depends on the electron density or, more specifically, the charges of all the other fragments
\begin{equation}\label{eq:fragment_hamiltonian}
H_{\mu \nu}^{X}=H_{\mu \nu}^{S C C, X}+V_{\mu \nu}^{X}
\end{equation}
where, $V_{\mu \nu}^{X}$, is the electrostatic potential (ESP) that acts on fragment $X$ and is defined as: 
\begin{equation}
    V_{\mu \nu}^{X} = \frac{1}{2} S_{\mu \nu}^X \sum_{K \notin I}^N \sum_{C \in K}^{N_K} \left(\gamma_{A C}+\gamma_{B C}\right) \Delta q_{C} \label{eq:electrostatic_potential}
\end{equation}
This makes it necessary to perform the SCC iterations for fragments self-consistently so that one iteration step per fragment is made to update all charges simultaneously. The obtained self-consistent charges are afterwards used for the calculation of the SCC iterations on pairs for which the ES-DIM approximation is not used. \\

Calculating the electronic ground state in the context of FMO(-DFTB) allows one to make accurate calculations of the energy, but one does not obtain MOs of the entire system. Instead, one obtains only the MOs for each of the fragments and all fragment pairs that are not affected by the ES-DIM approximation. 
Since they are obtained by independent calculations such MOs are not orthogonal. However, it is possible, as shown by Tsuneyuki \textit{et al.} \cite{tsuneyuki_molecular_2009}, to construct orthogonal MOs for the whole system and to transform the Hamiltonian into such basis by using Löwdin's approach\cite{lowdin_nonorthogonality_1950}
\begin{equation}\label{LCMO_fock_4}
    \mathbf{H}^{\prime}=\mathbf{S}^{-1 / 2} \mathbf{H}^{\mathrm{LCMO}} \mathbf{S}^{-1 / 2}.
\end{equation}
Since the computation of the matrix inverse of the total overlap is computationally demanding and the overlap matrix in the basis of the fragments and pairs is almost diagonal, we approximate it in first order by
\begin{equation}
    \mathbf{S}^{-1 / 2} \approx \frac{3}{2} \mathbf{1} - \frac{1}{2} \mathbf{S}
    \label{eq:invers_S_approx}
\end{equation}
where the overlap matrix matrix elements are given by
\begin{equation}
    S_{pq}^{IJ} = \langle \varphi_p^I | \varphi_q^J \rangle
\end{equation}
and $\varphi_p^I$ is the $p$-th MO of fragment $I$. 
The Hamiltonian of the whole system, $\mathbf{H}^{\mathrm{LCMO}}$, can be constructed from the Hamiltonian matrices of the corresponding fragments and pairs as
\begin{equation}\label{LCMO_fock_1}
    \mathbf{H}^{\mathrm{LCMO}}=\sum_{I}^N \oplus \mathbf{H}^{I}+ \frac{1}{2} \sum_{I}^N \sum_{J \neq I}^N \oplus\left(\mathbf{H}^{I J}-\mathbf{H}^{I} \oplus \mathbf{H}^{J}\right).
\end{equation}
The $\oplus$ sign indicates that each block in the total Hamiltonian matrix is filled with the Hamiltonian of the corresponding fragment ($\mathbf{H}^{I}$) or fragment pair ($\mathbf{H}^{IJ}$). 

In contrast to the diagonal Hamiltonian matrix of the fragments 
\begin{equation}\label{LCMO_fock_2}
  \epsilon_{a}^X    =  \delta_{a b} H^X_{a b},
\end{equation}
where $\epsilon_{a}$ is the $a$-th MO energy of fragment $X$, the Hamiltonian of pair $IJ$ accounts for the fact that the MOs between the fragments might not be strictly orthogonal
\begin{equation}\label{LCMO_fock_3}
    H^{IJ}_{ab} = \sum_{r} \epsilon_{r}^{IJ} \langle \varphi_a^I \mid \varphi_r^{IJ} \rangle \langle \varphi_b^J \mid \varphi_r^{IJ} \rangle 
\end{equation}
where the sum runs over all MOs of the pair $IJ$. Of course, it would now be possible to determine the eigenvectors of the Hamiltonian and thereby obtain the molecular orbitals of the entire system. In principle, once the MOs of the whole system are available, the excited states could be calculated using standard TD-DFTB procedure. While this seems straightforward, the disadvantage is that the calculation of the excited states are as costly as in a non-FMO approach and one would lose the scalability of the method. 

\subsection{Excited states in the frame of FMO-LC-DFTB}\label{sec:quasi_diabatic_states}
An alternative way to compute excited states in the framework of the FMO-Ansatz is motivated by the idea that an excited state wavefunction of the whole system can be expressed as a linear combination of basis states of the respective fragments and fragment pairs. Since we restrict ourselves to the interaction between two fragments in FMO, we also use only locally excited (LE) states on one fragment and charge-transfer (CT) states between two fragments as basis states. The electronic excited state wavefunction, $|\Psi\rangle$, of a large molecule assembly can then be expressed as 
\begin{equation}
\left|\Psi\right\rangle=\sum_{I}^N \sum_{m}^{N_{\mathrm{LE}}} c_I^m \left|\mathrm{LE}_I^m \right\rangle+ \sum_{I}^{N} \sum_{J \neq I}^{N} \sum_{m}^{N_{\mathrm{CT}}} c_{I\rightarrow J}^m \left| \mathrm{CT}_{I \to J}^{m}\right\rangle    
\label{eq:es_wf}
\end{equation}
where the coefficients for fragment LE and CT configuration state functions (CSFs) are given by $c_I^m$ and $c_{I\rightarrow J}^m$, respectively. The coefficients are obtained by solving the eigenvalue problem $\mathbf{Hc} = \mathbf{Ec}$. Notice that we assume that the basis states are orthogonal such that $\mathbf{S} = \mathbf{1}$. However, in the calculation of the Hamiltonian matrix elements the overlap matrix is explicitly included. This is in line with other semi-empirical methods that employ the same approximation. On the basis of FMO Hartree-Fock theory this approach was introduced by Fujita and Mochizuki\cite{fujita_development_2018}. However, we will show that our definition of the CT states differs and how this impacts not only the energies but also the scaling of the method. 
The ansatz in eq. \ref{eq:es_wf} allows one to choose the energetically lowest basis states by setting the number of LE states per fragment ($N_{\mathrm{LE}}$) and the number of CT states per pair ($N_{\mathrm{CT}}$) to appropriate values. The number of eigenvalues for the corresponding eigenvalue problem then is given by $N \cdot N_{\mathrm{LE}} + (N^2 - N) \cdot N_{\mathrm{CT}}$. Furthermore, since these basis states are per definition locally excited or charge-transfer states the adiabatic states' eigenvectors (cf. \ref{eq:es_wf}) contain the state character information. In principle, it would also be possible to extend this approach to doubly or even higher excited states.  \\
The LE states, $\left|\mathrm{LE}_{\mathrm{I}}^m\right\rangle$, are calculated as singly excited states of a fragment. We restrict ourselves in this work to singlet  states, but it should be noted that this approach can be extended to triplet excited states in a straightforward way. The $m$-th excited state (S$_m$) on fragment $I$ defines the following LE basis state: 
\begin{align}
    \left|\mathrm{LE}_{\mathrm{I}}^m\right\rangle &= \sum_{i \in I}\sum_{a \in I} T_{i a}^{m(I)}\frac{1}{2}\left(a_{\mathrm{Ia}, \alpha}^{\dagger} a_{\mathrm{Ii}, \alpha}+a_{\mathrm{Ia}, \beta}^{\dagger} a_{\mathrm{Ii}, \beta}\right)|G\rangle \\
    &= \sum_{i \in I}\sum_{a \in I}  T_{i a}^{m(I)} | \Phi_{I}^{i\to a} \rangle
    \label{eq:LE_states}
\end{align}
Here, $a_{I a, \alpha}^{\dagger}$ and $a_{I a, \beta}$ are the creation and annihilation operators of the $\alpha$ and $\beta$ spin electron in the $a$th orbital of fragment $I$, and $|G\rangle$ is the ground-state wavefunction.
$\mathbf{T}^{m(I)}$ is the one-particle transition density matrix of the $m$-th excited state of fragment $I$ in the MO basis and $|\Phi_{I}^{i\to a} \rangle$ is the configuration state function of the excitation from the occupied orbital $i$ to the virtual orbital $a$ on fragment $I$.\\
Similarly, the $m$-th intermolecular CT state from fragment $I$ (hole) to fragment $J$ (particle) can be defined as: 
\begin{align}
    \left|\mathrm{CT}_{I \to J}^{m}\right\rangle &=\sum_{i \in I}\sum_{a \in J} T_{i a}^{m(IJ)} \frac{1}{2}\left(a_{\mathrm{Ja}, \alpha}^{\dagger} a_{\mathrm{Ii}, \alpha}+a_{\mathrm{Ja}, \beta}^{\dagger} a_{\mathrm{Ii}, \beta}\right)|G\rangle \\
    &= \sum_{i \in I}\sum_{a \in J} T_{i a}^{m(IJ)} | \Phi_{I \to J}^{i\to a} \rangle
    \label{eq:CT_states}
\end{align}
The transition between the fragments $I$ and $J$ is restricted to the occupied orbitals of monomer $I$ and the virtual orbitals of monomer $J$. The definition of the CT state differs to the approach of Fujita and Mochizuki\cite{fujita_development_2018} as they use the following  single configuration state function as a CT state
\begin{equation}
    \left|\mathrm{CT}_{I \to J}^{i \to a}\right\rangle=\frac{1}{2}\left(a_{\mathrm{Ja}, \alpha}^{\dagger} a_{\mathrm{Ii}, \alpha}+a_{\mathrm{Ja}, \beta}^{\dagger} a_{\mathrm{Ii}, \beta}\right)|G\rangle =  | \Phi_{I \to J}^{i\to a} \rangle.
    \label{eq:CT_states_fujita}
\end{equation}
Both approaches have their merits, and we will briefly explain below why we define CT states as shown in equation \ref{eq:CT_states}. The latter necessitates solving the CIS matrix for the lowest $N_{\mathrm{CT}}$ charge-transfer states of each pair. Therefore, the construction of the basis states will no longer scale linearly with the number of fragments but quadratically. Nevertheless, this approach has the advantage that all electronic couplings between CT states on a pair are omitted since these are eigenstates of the pair. Furthermore, in both approaches, a restriction of the considered CT states or occupied and virtual orbitals is necessary. Using equation \ref{eq:CT_states}, we can choose to use the $N_{\mathrm{CT}}$ lowest energy CT states and ensure that all relevant orbitals for these states are considered, whereas with eq. \ref{eq:CT_states_fujita}, the selection of states is made by considering $N_{\mathrm{hole}}$ occupied and  $N_{\mathrm{particle}}$ virtual MOs. A higher number of included orbitals quickly leads to very large eigenvalue problems as the number of considered CT states scales with $(N^2 - N) \cdot N_{\mathrm{hole}} \cdot N_{\mathrm{particle}}$. We will show in section \ref{sec:accuracy_excited_states} how the accuracy of the excitation energies and the computational scaling depend on using either eq. \ref{eq:CT_states} or \ref{eq:CT_states_fujita} for the definition of charge-transfer states and that the usage of eq. \ref{eq:CT_states} is generally preferable. 

\subsection{Hamiltonian matrix elements}\label{sec:DiabaticHamiltonian}
The definition of basis states in eq. \ref{eq:LE_states} and \ref{eq:CT_states} allows to express the matrix elements of the system Hamiltonian. We will first focus on the excitation energies (diagonal matrix elements) of LE and CT states. The energies are calculated by a \dftb calculation of the fragment (pair)
 in the framework of linear-response LC-TD-DFT, which was adapted to tight-binding DFT by Niehaus \textit{et al.}\cite{niehaus_tight-binding_2001} and has been extended to include the LC correction by some of us\cite{humeniuk_long-range_2015}. In linear-response TD-DFT excited states are computed by solving the non-Hermitian eigenvalue problem 
\begin{equation}
    \left(\begin{array}{ll}
\mathbf{A} & \mathbf{B} \\
\mathbf{B} & \mathbf{A}
\end{array}\right)\left(\begin{array}{l}
\mathbf{X} \\
\mathbf{Y}
\end{array}\right)=\bm{\omega}\left(\begin{array}{cc}
\mathbf{1} & 0 \\
0 & -\mathbf{1}
\end{array}\right)\left(\begin{array}{l}
\mathbf{X} \\
\mathbf{Y}.
\end{array}\right)\label{eq:Casida}
\end{equation}
In the framework of \tddftb the matrices $\mathbf{A}$ and $\mathbf{B}$ are defined as\cite{humeniuk_long-range_2015}
\begin{align}
    A_{i a, j b} &=\delta_{i j} \delta_{a b}\left(\epsilon_a-\epsilon_i\right)+2 \sum_{A}\sum_{B} q^{ia}_A \gamma_{AB} q_B^{jb} - \sum_{A} \sum_{B} q^{ij}_A \gamma_{AB}^{\mathrm{lr}} q^{ab}_B \\
    B_{i a, j b} &= 2 \sum_{A}\sum_{B} q^{ia}_A \gamma_{AB} q_B^{jb} - \sum_{A} \sum_{B} q^{ib}_A \gamma_{AB}^{\mathrm{lr}} q^{aj}_B,
\end{align}
where $\epsilon_i$ is the energy of the $i$-th molecular orbital. The orbital energies $\epsilon_i$ are represented by the diagonal elements of the orthogonalized total Hamiltonian $H^\prime_{ii}$. Using the Tamm-Dancoff (TDA) approximation ($\mathbf{B} = 0$), eq. (\ref{eq:Casida}) is reduced to a simple Hermitian eigenvalue problem
\begin{equation}
    \mathbf{A} \mathbf{X} = \bm{\omega} \mathbf{X},
\end{equation}
which can be solved iteratively for the lowest eigenvalues by employing the algorithm developed by Davidson\cite{davidson_iterative_1975}. This is in principle equivalent to the configuration interaction singles (CIS) procedure. The energies of the CT states are calculated in analogy to the local excitations, where the excitations are restricted to the transition from the occupied orbitals of one monomer to the virtual orbitals of the other one.

To calculate the off-diagonal matrix elements of the Hamiltonian, the couplings between the basis states are required.
In accordance with Ref. \cite{fujita_development_2018}, the Hamiltonian can be split into one electron
\begin{equation}\label{One_electron_hamiltonian}
    \langle \Phi_{I \to J}^{i\to a}|H_{1e}|\Phi_{K \to L}^{j \to b} \rangle = \delta_{I K} \delta_{i j} H_{a b}^{\prime}-\delta_{J L } \delta_{a b} H_{i j}^{\prime}
\end{equation}
and two-electron contributions
\begin{equation}\label{two_electron_hamiltonian}
    \langle \Phi_{I \to J}^{i\to a} | H_{2 e} |\Phi_{K \to L}^{j \to b} \rangle =2\left(i^{(I)} a^{(J)} \mid j^{(K)} b^{(L)}\right)-\left(i^{(I)} j^{(K)} \mid a^{(J)} b^{(L)}\right),
\end{equation}
which were derived using the usual Slater-Condon rules. The two-electron coupling matrix elements of the Hamiltonian require the calculation of the four-center electron repulsion integrals (\textit{cf}. eq. \ref{two_electron_hamiltonian}). In contrast to the work of Fujita and Mochizuki\cite{fujita_development_2018}, the off-diagonal elements between the basis states that involve three or four fragments are not neglected.

\subsubsection{LE-LE matrix elements}
 The coupling between locally excited states is given only by the two-electron part, as the one electron coupling between locally excited states vanish due to the Slater-Condon rule,
\begin{align*}
\left\langle \mathrm{LE}_{I}^{m}\left|H \right| \mathrm{LE}_{J }^{n}\right\rangle &=\sum_{ia \in I} \sum_{jb \in J} T_{i a}^{m(I)} T_{jb}^{n(J)}[2(i a \mid j b) \\
 &-(ij \mid ab)] \label{le_le_coulomb} \numberthis{}{}  \\
\approx& 2 \sum_{A\in I} \sum_{B \in J} q_{\mathrm{tr}, A}^{m(I)} \gamma_{AB} q_{\mathrm{tr}, B}^{n(J)}\\
-& \sum_{A\in IJ} \sum_{B \in IJ} 
\sum_{ia \in I} \sum_{jb \in J} T_{i a}^{(\mathrm{Im})} T_{jb}^{(J n)} q_{A}^{ij} \gamma_{AB}^{\mathrm{lr}} q_{B}^{ab}, \numberthis{}{} \label{le_le_coupling}
\end{align*}
where
\begin{equation}
    q_{\mathrm{tr}, A}^{m(I)} = \sum_{ia} q_{A}^{ia} T_{i a}^{m(I)} 
\end{equation}
is the transition charge of the $m$-th excited state of the fragment $I$ on atom $A$. $q_{A}^{ij}$ and $q_{B}^{ab}$ are the atom-centered transition charges between the occupied and virtual orbitals of the two monomer fragements $I$ and $J$ (\textit{cf}. eq. (\ref{eq:q_between_mos})). The first term of the LE-LE coupling (\textit{cf}. eq. \ref{le_le_coulomb}) characterizes the Coulomb interaction between the transition densities of the respective excitations. The second term represents the exchange interaction.
In the case of LE states on far separated fragments, fragment pairs for which the ES-DIM approximation is used, the exchange contribution can be neglected as the overlap between both states will be zero: 
\begin{equation}
    \left\langle \mathrm{LE}_{I}^{m}\left|H\right| \mathrm{LE}_{J }^{n}\right\rangle \approx 2 \sum_{A\in I} \sum_{B \in J} q_{\mathrm{tr}, A}^{m(I)} \gamma_{AB} q_{\mathrm{tr}, B}^{n(J)}
\end{equation}
Therefore, only the long-range Coulomb interaction is taken into account in this case. 

\subsubsection{LE-CT matrix elements}
The Hamiltonian matrix elements between an LE state on fragment $I$ and a CT state which is formed by linear combination of holes on fragment $J$ and electrons on fragment $K$ contains a one-electron part
\begin{align*} 
    \left\langle\mathrm{\mathrm{LE}}_{I}^{m}\left|H_{1 e}\right| \mathrm{CT}_{J \to K}^{n}\right\rangle = & \delta_{I J} \sum_{ia \in I}\sum_{b \in K} T_{ia}^{m(I)} T_{ib}^{n(IK)} H_{a b}^{\prime} \\ &-\delta_{I K}
    \sum_{ia \in I}\sum_{j \in J} T_{i a}^{m(I)}T_{ja}^{n(JI)} H_{i j}^{\prime}
    \label{eq:LE_CT_1e_coupling_2} \numberthis{}{}
\end{align*}
as well as a contribution from the two-electron interaction
\begin{align*}
\left\langle \mathrm{LE}_{\mathrm{I}}^m\left|H_{2 e}\right| \mathrm{CT}_{J \to K}^{n}\right\rangle &= \sum_{ia \in I} \sum_{j \in J} \sum_{b \in K}  T_{i a}^{m(I)} T_{j b}^{n(JK)} [2(i a \mid jb ) \\ &-( i j \mid ab )] \numberthis{}{} \\
\approx 2\sum_{A \in I}\sum_{B \in JK}& q_{\mathrm{tr}, A}^{m(I)} \gamma_{AB} q_{\mathrm{tr}, B}^{n(JK)} \\
- \sum_{A \in IJ}\sum_{B \in IK
} &\sum_{ia \in I}  \sum_{j \in J}\sum_{b \in K}T_{ia}^{m(I)} T_{jb}^{n(JK)} q_{A}^{ij} \gamma_{AB}^{\mathrm{lr}} q_{B}^{ab}. \label{le_ct_coupling} \numberthis{}{}
\end{align*}
The first and second term in eq. \ref{le_ct_coupling} correspond to the Coulomb and exchange interaction between the transition densities of the LE and CT state. The one-electron contribution is only non-zero if the CT state shares one of its fragment with the LE state and if the other fragment is in spatial proximity, so that the ES-DIM approximation is not used, because otherwise the matrix element $H_{pq}^{\prime}$ will be zero. If both fragments of the CT state are far apart (ES-DIM approx.) from the fragment of the LE state then the exchange term of eq. \ref{le_ct_coupling} will be zero and is neglected. In this case the matrix element becomes
\begin{equation}
    \left\langle \mathrm{LE}_{\mathrm{I}}^m\left|H\right| \mathrm{CT}_{J \to K}^{n}\right\rangle  
\approx 2\sum_{A \in I}\sum_{B \in JK} q_{\mathrm{tr}, A}^{m(I)} \gamma_{AB} q_{\mathrm{tr}, B}^{n(JK)}
\end{equation}
and only the Coulomb interaction is calculated.

\subsubsection{CT-CT matrix elements}
The one-electron CT--CT coupling vanishes for $I \neq K$ or $K \neq L$ (\textit{cf}. eq. \ref{One_electron_hamiltonian}), and thus, is reduced to the diagonal contributions, which are included in the LC-TD-DFTB calculation of the CT state. 
The two-electron off-diagonal coupling between two charge-transfer states is given by
\begin{align*}
&\left\langle \mathrm{CT}_{I \to J}^{m(IJ)}\left|H \right| \mathrm{CT}_{K \to L}^{n (KL)}\right\rangle = \\ 
&\sum_{i \in I}\sum_{a \in J}\sum_{j \in K}\sum_{b \in L} T_{ia}^{m(IJ)} T_{jb}^{n(KL)}  [2(i a \mid j b) -(i j \mid a b)] \numberthis{}{} \\
&\approx 2 \sum_{A\in IJ} \sum_{B \in KL} q_{\mathrm{tr}, A}^{m(IJ)} \gamma_{AB} q_{\mathrm{tr}, B}^{n(KL)}\\
&- \sum_{i \in I}\sum_{a \in J}\sum_{j \in K}\sum_{b \in L} \sum_{A \in IK} \sum_{B \in JL} T_{ia}^{m(IJ)} T_{jb}^{n(KL)}q_{A}^{ij} \gamma_{AB}^{\mathrm{lr}} q_{B}^{ab} \numberthis{}{} \label{ct_ct_coupling}
\end{align*}
As shown for the LE-LE and LE-CT matrix elements, we also neglect the exchange part of the coupling for the CT-CT couplings in case if either fragment $I$ and $K$ or $J$ and $L$ are different fragments that are far apart. Thus in this case the matrix element simplifies to
\begin{equation}
    \left\langle \mathrm{CT}_{I \to J}^{m(IJ)}\left|H\right| \mathrm{CT}_{K \to L}^{n (KL)}\right\rangle 
\approx 2 \sum_{A\in IJ} \sum_{B \in KL} q_{\mathrm{tr}, A}^{m(IJ)} \gamma_{AB} q_{\mathrm{tr}, B}^{n(KL)} \label{eq:CT_CT_ESDIM}
\end{equation}

\section{Computational procedure}\label{sec:computational_procedure}
\begin{figure}[tbh!]
    \centering
    \includegraphics[width=\linewidth]{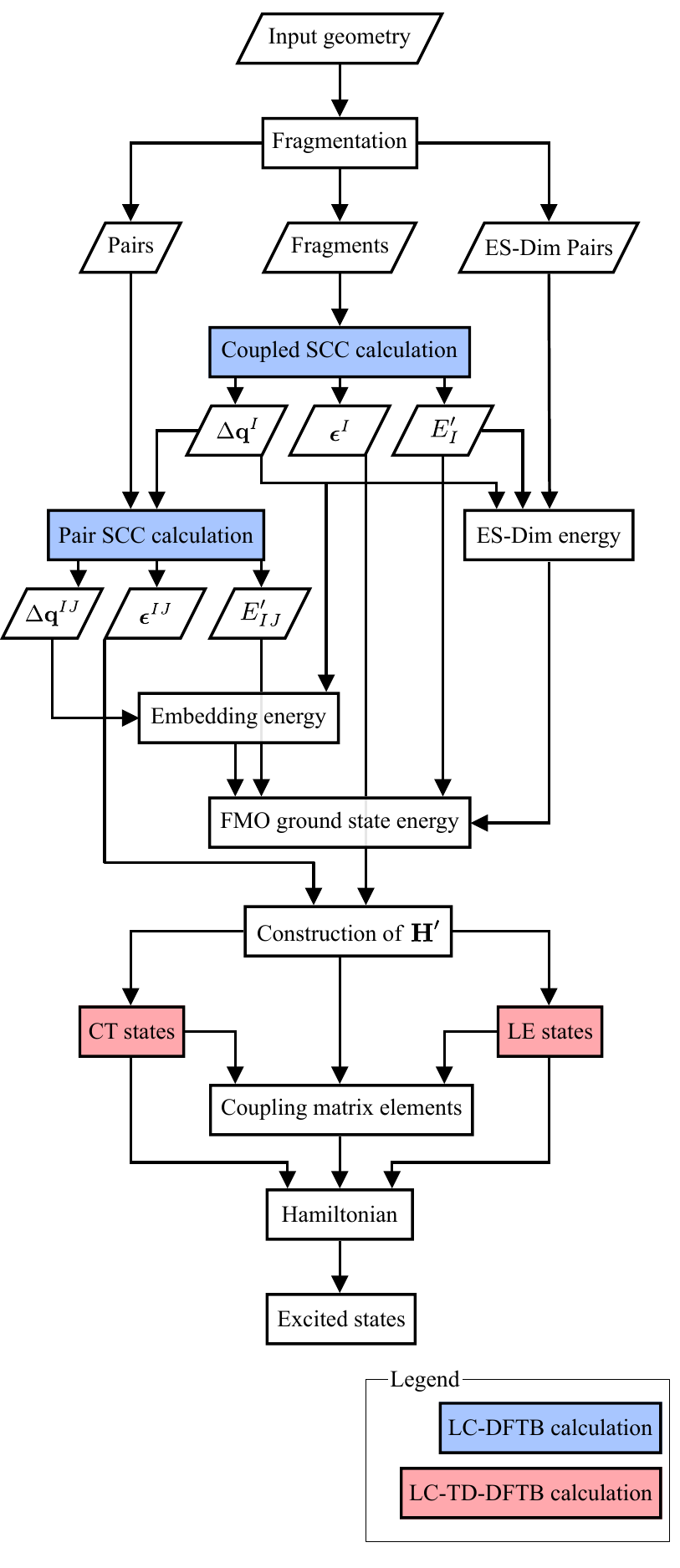}
    \caption{Flowchart of the computational procedure of the entire \fmo calculation.}
    \label{flowchart}
\end{figure}
Herein, the sequence of the steps necessary to carry out a \fmo calculation is briefly summarized. The flowchart displayed in Fig. \ref{flowchart} shows the 
order of the different steps of a \fmo calculation and how the various parts of the calculation are connected. 

In the first step, the input geometry is partitioned into the different monomer, pair and ES-DIM pair fragments. The ES-DIM approximation\cite{nakano_fragment_2002} is applied if the closest distance between two fragments exceeds the threshold value of twice the sum of the van der Waals radii of the closest atoms. 

Subsequently, the coupled lc-DFTB SCC iterations of the monomer fragments are performed until the convergence of all monomer calculations is achieved. As the electrostatic potential between all monomer fragments is required to calculate the monomer Hamiltonian (\textit{cf.} eq. \ref{eq:fragment_hamiltonian}), the SCC iterations for all fragments are computed simultaneously in order to update the charges of all monomers in each step. The final self-consistent charges are stored, as they are necessary for the calculation of the pair, ES-DIM pair and embedding energies. The Anderson acceleration is used to speed up the convergence of the SCC routine \cite{anderson_iterative_1965,zhang_globally_2020}. 

Then, using the electrostatic potential between all fragments (\textit{cf}. eq. \ref{eq:electrostatic_potential}) obtained from the previous step, the pair energies are calculated. The charges of the pair fragments are stored so that they can be used in the calculation of the embedding energy. The ES-DIM pair energy (\textit{cf}. eq. \ref{eq:esdim_energy}) and the embedding energy (\textit{cf}. eq. \ref{eq:embedding_energy}) are determined, using the charges of the monomer and pair fragments obtained from the previous SCC routines. The results of the fragment calculations are combined according to eq. (\ref{eq:total_fmo_energy}), which yields the ground-state energy  of the \fmo approach.

Thereafter, the LCMO-Hamiltonian matrix is built according to eq. \ref{LCMO_fock_4}--\ref{LCMO_fock_3}, using the orbital energies, orbital coefficients and overlap matrices of the fragments. The $\mathbf{H'}$ matrix is obtained by subsequent Löwdin orthogonalization. In order to construct the LE and CT basis states, \tddftb excited state calculations are performed for both the monomers, the pairs and the ES-DIM pairs, where the orbital energies of the respective fragments are replaced by the diagonal matrix elements of the $\mathbf{H'}$ matrix. 

Subsequently, the off-diagonal matrix elements of the Hamiltonian are calculated according to the expressions for the couplings between the LE and CT states  given in eq. \ref{le_le_coupling} -- \ref{eq:CT_CT_ESDIM}. 
As the calculation of the exchange terms in the eq. (\ref{le_le_coupling}), (\ref{le_ct_coupling}) and (\ref{ct_ct_coupling}) includes all possible transitions between the orbitals of the monomer pairs, a screening of the one-particle transition density matrices $\mathbf{T}$ was introduced, where only matrix elements over a certain threshold are considered. Thus, the number of transitions is limited to the actually contributing occupied and virtual orbitals of the fragments. 

At the end of the \fmo calculation, the Davidson diagonalization is utilized to obtain the excited state energies and coefficients, which are then used to calculate the oscillator strengths.

The \tddftb and \fmo methods were implemented in our own software package DIALECT, which is available on Github\cite{noauthor_dialect_2022}. 

All calculations regarding the benchmarks and scaling tests were performed on a single core of a computing node with dual E5-2680 Xeon CPUs (2.4 GHz, 14 cores each), 188 Gb of DDR4 memory and a SATA hard drive. 

The DFTB parameter set ob2-split\cite{vuong_parametrization_2018} was employed in all calculations. In the case of the benchmark calculations, a long-range radius of 3.03~$a_0$ was used. In order to compensate the overestimation of the charge-transfer energies in the pentacene clusters, IP-EA tuning\cite{stein_reliable_2009} was employed and an optimized long-range radius was used for these systems. This procedure will be described in greater detail in section \ref{ch:pentacene_application}.

The Mercury\cite{macrae_it_2020} program was employed to generate the various clusters of the anthracene, pentacene and perylen bisimide systems from their respective crystal structures\cite{mason_crystallography_1964,holmes_nature_1999,lin_halochromic_2011}.

In case of the scans of potential energy curves of the pyrene dimer, the DFT-D3 dispersion correction\cite{grimme_consistent_2010,grimme_effect_2011} was calculated using the simple-dftd3\cite{ehlert_simple-dftd3_2021} program and the ob2-split dispersion parameters\cite{vuong_parametrization_2018}.
\begin{figure*}[tb]
    \centering
    \includegraphics[width=\linewidth]{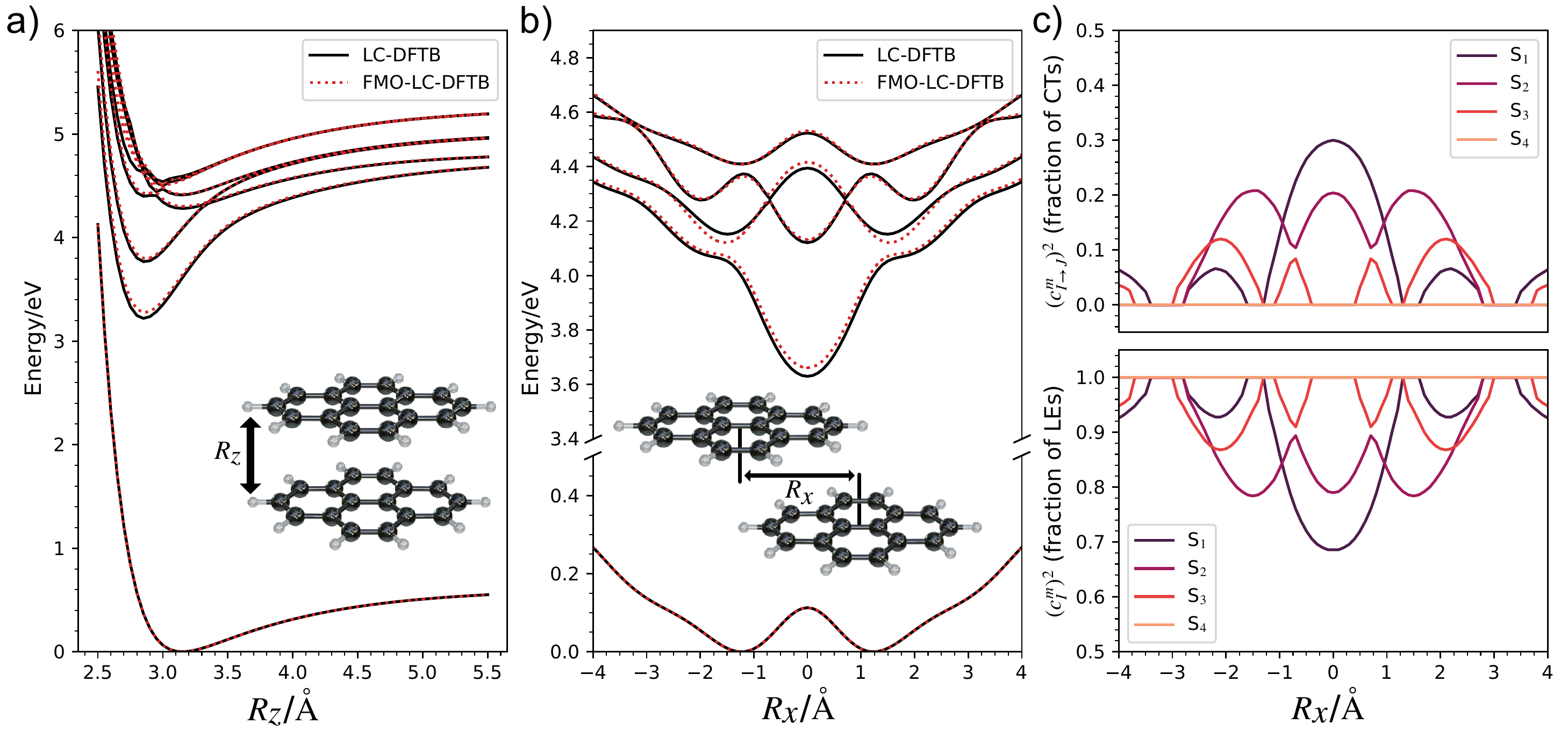}
    \caption{(a) Potential energy curves of the ground state and the first 6 excited states of the pyren dimer along the $\pi$--$\pi$ stacking coordinate $R_z$. (b) Potential energy curves of the ground state and the first 4 excited states along the parallel shift coordinate $R_x$ at a $\pi$--$\pi$ stacking distance of 3.1 \AA{}. (c) The contributions of the locally excited and charge transfer states to the excited states shown in b) for different values of $R_x$. For each monomer 20 local excitations and for each pair 15 charge-transfer states were used.}
    \label{dimer_scan}
\end{figure*}
\section{Results}\label{sec:results}
\subsection{Accuracy of \fmo excited states}\label{sec:accuracy_excited_states}
The accuracy of the \fmo methodology is evaluated calculating the excitation energies of $\pi$-stacked dimers
of pyrene. The potential energy curves of the first 6 excited states were calculated  as a function of the $\pi$--$\pi$ stacking distance $R_z$ and the parallel shift coordinate $R_x$. As a reference we performed calculations with the conventional \tddftb approach. The main purpose of this comparison is to test the FMO approach and the validity of the employed approximations and not to benchmark the general performance. The latter has been thoroughly benchmarked in previous works.\cite{vuong_parametrization_2018, bold_benchmark_2020, fihey_performances_2019} In the case of the \fmo calculation, 20 locally excited states for each monomer and 15 charge-transfer states for each pair were included. In Fig. \ref{dimer_scan}a, the electronic state energies of the \fmo and \tddftb methods are shown for a scan of the $\pi$--$\pi$ stacking distance $R_z$. While there is a slight deviation from the \tddftb energies at distances from 2.75~\AA{} to 3.0~\AA{}, the deviation between the two methods decreases with increasing interfragment distances. From a distance of 3.5~\AA{} onwards, the results of the \fmo calculation show an excellent agreement with the \tddftb energies. Below distances of 2.75~\AA{}, which already belong to the repulsive part of the potential energy curves, the deviation from the reference \tddftb calculation results from the non-orthogonality of the MOs.

Fig. \ref{dimer_scan}b shows the potential energy curves of the pyrene dimer for a scan of the parallel shift coordinate $R_x$ at a $\pi$--$\pi$ stacking distance of 3.1~\AA{}. Here, the results of the \fmo calculation are in good agreement with the \tddftb reference. 

As the excited state Hamiltonian is constructed using LEs and CTs as basis states, the analysis of the excited state composition becomes easily accessible. The contributions of the LE and CT states to the first 4 excited states for different parallel shift distances are shown in Fig. \ref{dimer_scan}c. As the shift along $R_x$ decreases, the contribution of the charge-transfer states increases. While the S$_1$, S$_2$ and S$_3$ state show CT character of up to 30\%, 20\% and 12\%, the $S_4$ onlyconsists of local excitations for all $R_x$ values.

In addition, the influence of the number of basis states on the accuracy of the excited states energies was investigated. To this end, the mean absolute errors (MAEs) of the first 6 excited states were calculated for different numbers of LE and CT states. Table \ref{Tab:MAE_pyrene_scan} shows the MAEs of the molecular aggregate for $\pi$--$\pi$ stacking distances of 2.5, 2.75, 3.0, 3.5, 4.0 and 5.0 \AA{}. The results confirm that the accuracy of the excited state energies of the dimer calculated at the \fmo level is satisfactory. At an intermolecular distance of 4.0 \AA{} the errors are as small as 2.7--5.6~meV. At 3.5~\AA{}, the MAEs are 4.8--9.9 meV, showing a an almost perfect agreement to the \tddftb energies.

\begin{table}[tbh]
\centering
\caption{Mean absolute errors (MAEs) of the first 6 excited states for the pyrene dimer at various interplanar distances $R_z$ and number of basis states.}
\label{Tab:MAE_pyrene_scan}
\setlength\tabcolsep{6.5pt}
\begin{tabular}{@{}cc|rrrrrr@{}}
\toprule
     &      & \multicolumn{6}{c}{MAE / meV}        \\ \midrule
N$_{\mathrm{LE}}$ & N$_{\mathrm{CT}}$ & 2.5 \AA{}& 2.75 \AA{} & 3.0 \AA{}  & 3.5 \AA{}& 4.0 \AA{}& 5.0 \AA{}\\ \midrule
5    & 5    & 433.7 & 136.4& 34.5  & 9.9 & 5.6 & 2.5 \\
10   & 5    & 406.8 & 118.2& 23.7 & 5.9 & 3.2 & 1.4 \\
10   & 10   & 192.9 & 74.2 & 21.6  & 5.6 & 3.2 & 1.4 \\
15   & 10   & 194.6 & 73.7 & 21.6  & 5.6 & 3.2 & 1.4 \\
20   & 15   & 186.2 & 72.7& 21.9  & 5.6 & 3.2 & 1.4 \\
30   & 20   & 178.4 & 71.0& 20.6  & 4.8 & 2.7 & 1.2 \\ \bottomrule
\end{tabular}
\end{table}
As expected, increasing the number of basis states improves the accuracy of the excited states energies. However, Table \ref{Tab:MAE_pyrene_scan} shows that a further increase from 20 LE and 15 CT to 30 LE and 20 CT states yields only a marginal improvement of the MAEs as the additional states have a negligible electronic coupling to the first 6 excited states.
At all physically relevant intermolecular separations, the \fmo method almost perfectly reproduces the PES for the pyrene dimer. The deviations start to be visible at distances around 3~\AA{}. Below 2.75~\AA{}, which is already in the strongly repulsive part of the potential, the non-orthogonality of the MOs causes the deviation from the results of the \tddftb reference as the error in the couplings between the LE and CT states increases. Nevertheless, at distances at around 2.75~\AA{} the MAEs are still relatively small; thus, the model should be sufficient for most physically realistic situations.

\begin{figure}[h!]
    \centering
    \includegraphics[width=\linewidth]{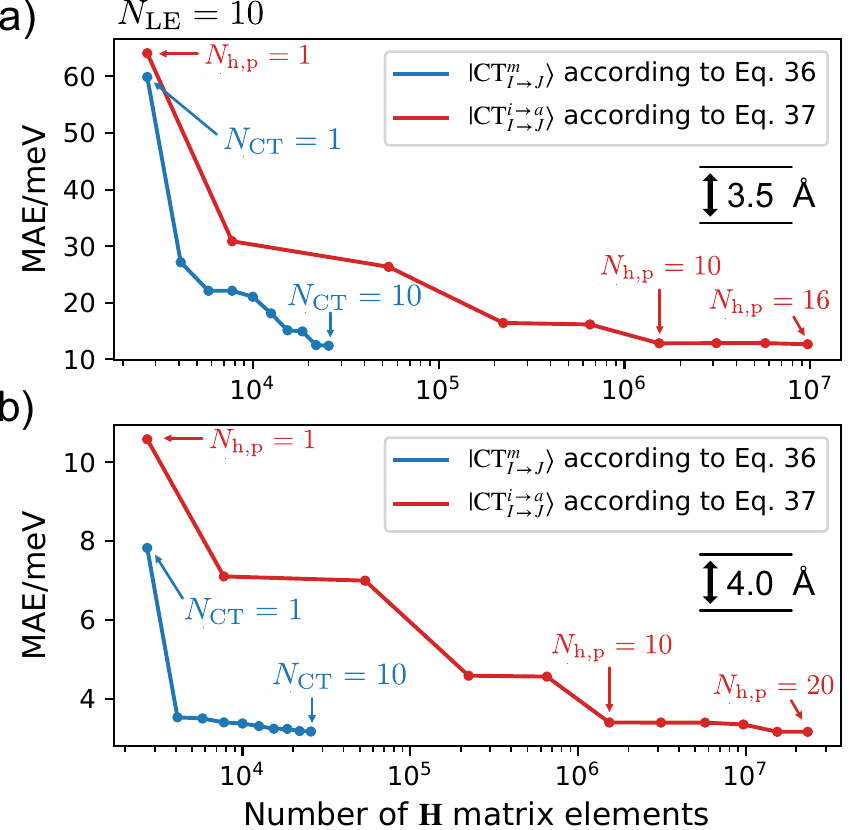}
    \caption{Comparison of the MAEs of both implementations of the charge-transfer states according to eq. (\ref{eq:CT_states}) and (\ref{eq:CT_states_fujita}) for different amounts of CT states at $\pi$--$\pi$ stacking distances of 3.5 and 4.0~\AA{} for a system of four stacked pyrene molecules.}
    \label{ct_comparison}
\end{figure}

To evaluate the accuracy of our implementation of the charge-transfer states (\textit{cf.} eq. \ref{eq:CT_states}) in comparison to the approach of Fujita \textit{et al.}\cite{fujita_development_2018} (\textit{cf.} eq. \ref{eq:CT_states_fujita}), the mean absolute errors of the first 40 excited states from the \tddftb reference of a $\pi$-stacked system of four pyrene monomers were calculated within both implementations. Fig. \ref{ct_comparison} shows the results for stacking distances of 3.5 and 4.0 \AA{}. For both methods, the number of LE states was set to 10 and the number of CT states was varied. As explained in section \ref{sec:quasi_diabatic_states}, in our implementation, the dimension of the Hamiltonian grows linearly as we consider the lowest $N_{\mathrm{CT}}$ states, and in the case of of Fujita's and Mochizuki's approach, the dimension of the Hamiltonian grows quadratically as all transitions between $N_{\mathrm{hole}}$ occupied and $N_{\mathrm{particle}}$ virtual orbitals are considered as separate CT states. As the amount of the charge-transfer states is increased for both implementations, the MAEs decrease in value. While both approaches achieve approximately the same accuracy, our method necessitates a much smaller size of the Hamiltonian. 

\begin{figure}[bth!]
    \centering
    \includegraphics[width=\linewidth]{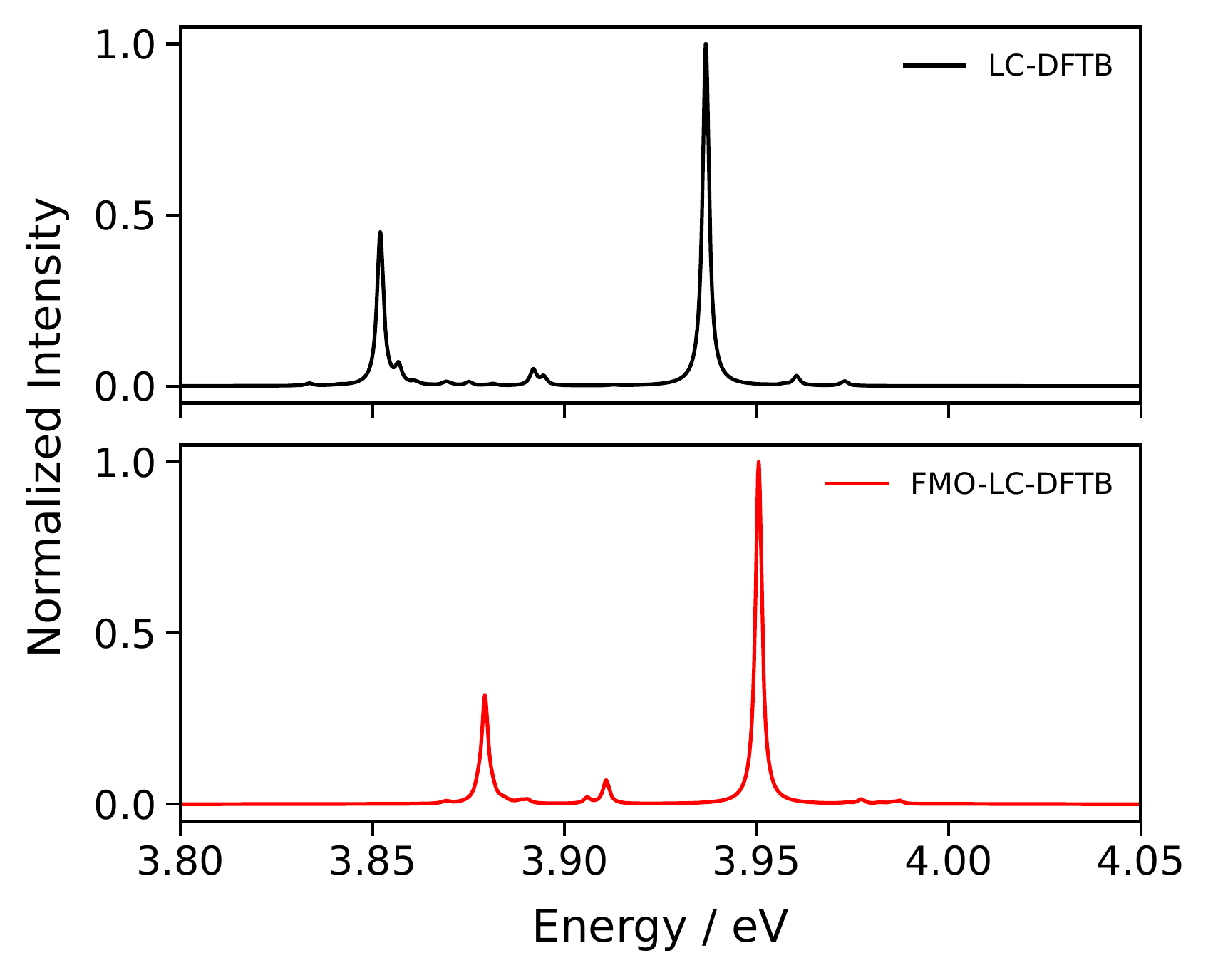}
    \caption{Simulated absorption spectra of an anthracene cluster containing 48 fragments (1152 atoms). The transitions to the first 40 excited states of the system were convolved using a Lorentzian line shape with a broadening of 2~meV. Two LE states and one CT state were used for each monomer and pair, respectively.}
    \label{spectrum_tda_vs_fmo}
\end{figure}

In order to investigate the accuracy of the \fmo method regarding the energies and oscillator strengths of larger systems, we compared the absorption spectra of an anthracene cluster composed of 48 fragments (1152 atoms) including the first 40 excited states for both theoretical methods. The basis was constructed using two LE states for each monomer and one CT state for each pair. The individual vertical transitions were convolved using a Lorentzian line-shape function with a width of 2~meV. The comparison between the calculated spectra is depicted in Fig. \ref{spectrum_tda_vs_fmo} and shows a slight blue shift of the \fmo energies by about 20--30 meV. However, since the general error in TD-DFT and, in particular, in \tddftb excitation energies is on the order of 0.5~eV\cite{vuong_parametrization_2018, fihey_performances_2019}, the additional error due to the FMO approach is hardly significant. Aside from slightly underestimating the intensity of the first peak, our approach could sufficiently reproduce the anthracene cluster's \tddftb absorption spectrum.

\subsection{Computational scaling of \fmo}
\begin{figure}[tbh!]
    \centering
    \includegraphics[width=\linewidth]{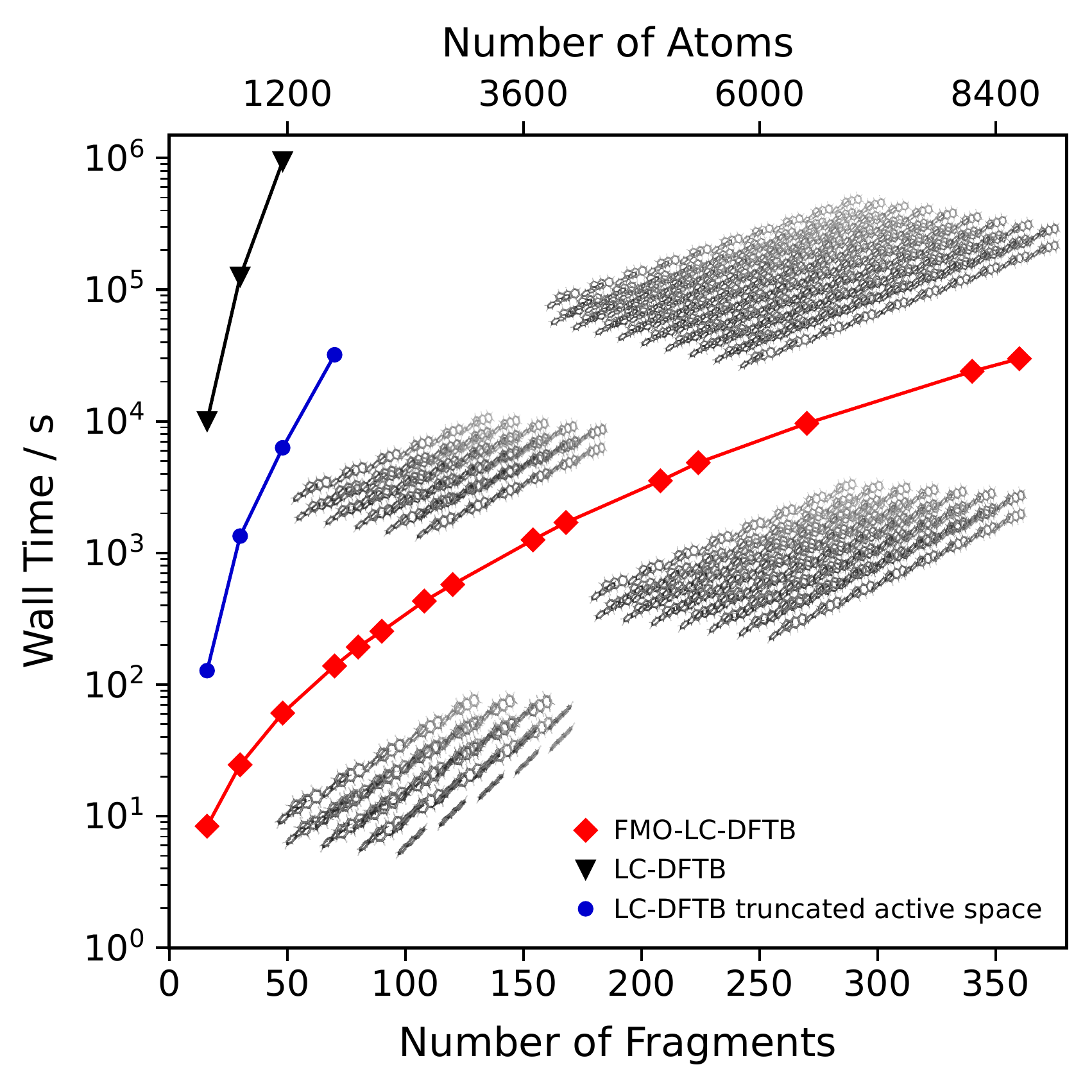}
    \caption{Wall time comparison for an excited state calculation of anthracene clusters of different size between \fmo and \tddftb. The difference between using all molecular orbitals for the active and a reduced active space of only considering 20~\% of the MOs is also shown.}
    \label{wall_time_fmo_vs_dftb}
\end{figure}

To investigate the computational efficiency of the developed method for the calculation of excited states, we compare the wall times of the method against conventional \tddftb for anthracene clusters containing up to 360 fragments (8640 atoms). In the case of the \tddftb approach, the number of fragments was limited to 48 (1152 atoms) due to the rapid increase in computational demand. For each cluster, the first 40 excited singlet states were calculated, and as in the previous anthracene calculation (\textit{cf.} Fig. \ref{spectrum_tda_vs_fmo}) two LE states per monomer and one CT state per pair were used for \fmo computations.
The wall times of the \fmo and \tddftb calculations are compared in Fig. \ref{wall_time_fmo_vs_dftb}. In the case of the anthracene cluster containing 48 fragments (1152 atoms), a speedup factor of approx. $1.5\times 10^4$ was achieved. Whereas the \tddftb calculation, using a full active space, took almost 11 days, the fragment approach finished in 60~seconds. 

\begin{figure}[bth!]
    \centering
    \includegraphics[width=\linewidth]{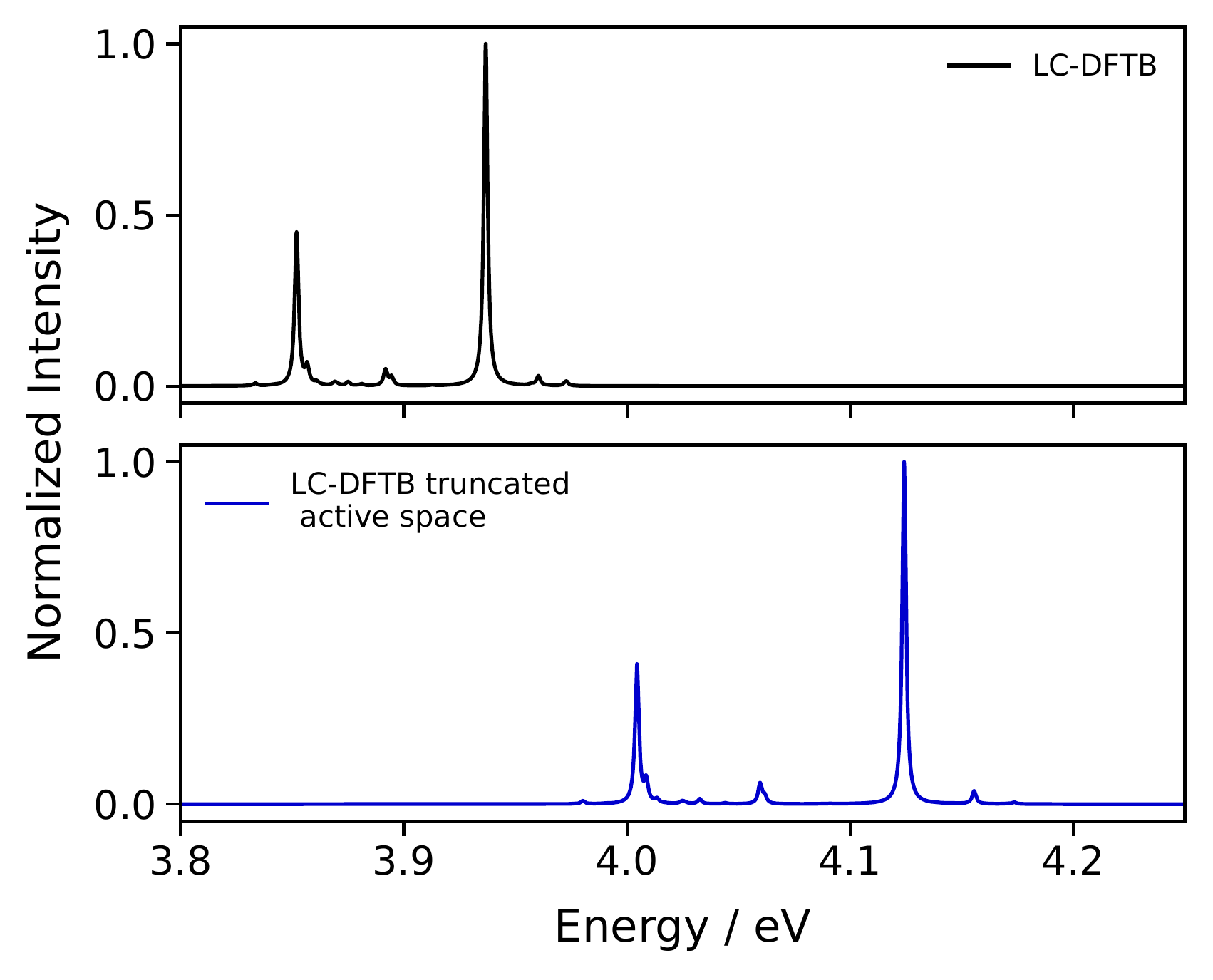}
    \caption{Simulated absorption spectrum (\tddftb: black, \tddftb with 20~\% active space: blue) of an anthracene cluster containing 48 fragments (1152 atoms). The transition to the first 40 excited states of the system were convolved using a Lorentzian line shape with a broadening of 2~meV.}
    \label{spectrum_tda_vs_truncated_tda}
\end{figure}

Furthermore, we analyzed the timings of the \tddftb method using only 20~\% of the active orbital space. As expected, this reduction in size of the $\mathbf{A}$ matrix in the TDA-DFTB approach significantly accelerates the computation of excited states for the anthracene cluster consisting of 48 fragments which took 6293~seconds. However, the \fmo approach is still around 100 times faster for this system and the speedup grows to factor of more than 200 for the next bigger anthracene cluster containing 70 fragments (1680~atoms). 
It should be noted that a restriction of the active spaces only lowers the prefactor of the method and not scaling. In addition, the restriction of the active orbitals leads to a loss of accuracy in (LC)-TD-DFTB as shown in Fig. \ref{spectrum_tda_vs_truncated_tda}. While the energies and oscillator strength of \fmo with only two LE states and one CT state has shown a good agreement with \tddftb, the reduction of the active orbital space to 20~\% leads to significantly larger errors of around 150~meV.

\begin{figure*}[bth!]
    \centering
    \includegraphics[width=\linewidth]{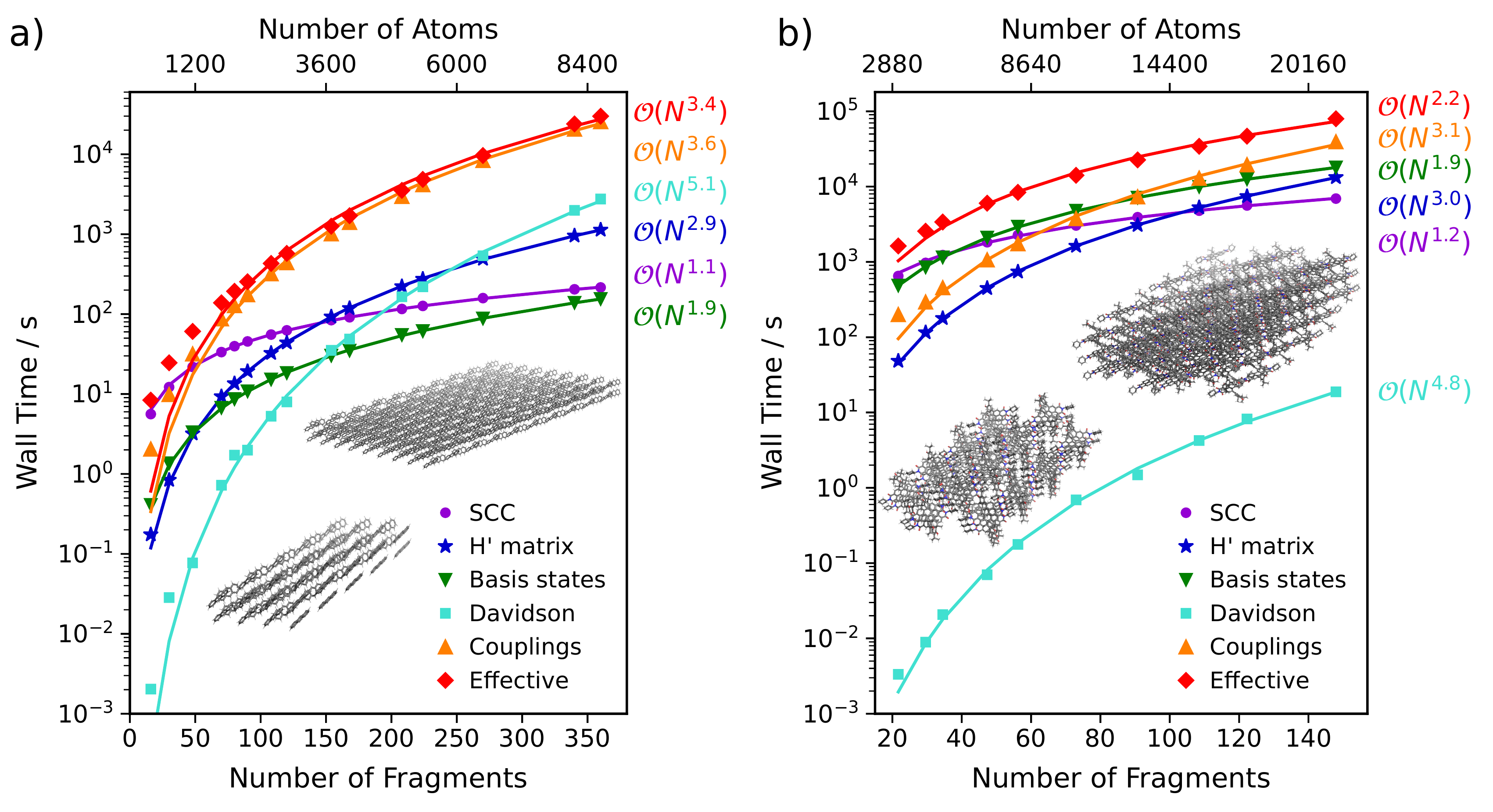}
    \caption{Timings of the \fmo excited state calculations for various a) anthracene and b) perylene bisimide clusters. The runtime scaling in relation to the system size is shown for different parts of the calculation on the right hand side of each figure. }
    \label{log_scaling}
\end{figure*}

In order to gain an insight into the time requirements of the different steps involved in a \fmo calculation, the total wall times have been split into separate parts. Fig. \ref{log_scaling}a shows the timings of five steps: 
\begin{enumerate}
    \item SCC: ground-state calculation of the entire systems
    \item $\bm{H^{\prime}}$ matrix: construction and Löwdin orthogonalization of the Hamiltonian matrix according to eq. \ref{LCMO_fock_4} --\ref{LCMO_fock_3}
    \item Basis states: calculation of LE and CT basis states of individual fragments and pairs according to section \ref{sec:DiabaticHamiltonian}
    \item Couplings: computation of one- and two-electron coupling matrix elements between all LE and CT states according to eq. \ref{le_le_coupling} -- \ref{eq:CT_CT_ESDIM}
    \item Davidson: diagonalization of the Hamiltonian using the iterative Davidson procedure. 
\end{enumerate}

In addition, the asymptotic scaling of the respective parts of the calculation was evaluated. With the increase in the size of the anthracene clusters, the Hamiltonian size grows up to a dimension of $129960^2$ for the system of 360 fragments (8640~atoms). Thus, the most time demanding steps are the calculation of the electronic couplings between the basis states and the following diagonalization for the lowest eigenvalues, which scale with factors of $\mathcal{O}(N^{3.6})$ and $\mathcal{O}(N^{5.1})$, respectively. The SCC routine implemented in our code shows the same nearly linear scaling of $\mathcal{O}(N^{1.1})$, which has been reported previously\cite{nishimoto_density-functional_2014}.
As the construction of the Fock matrix is limited by the scaling of the matrix multiplications during the Löwdin orthogonalization, it scales with a factor of $\mathcal{O}(N^{2.9})$. Compared to the other parts of the calculation, the scaling of the construction of the basis states is negligible with a factor of $\mathcal{O}(N^{1.9})$. 
The combination of all steps of the procedure results in an effective scaling of $\mathcal{O}(N^{3.4})$ for the total run time.

A molecular system involving larger fragments was investigated to verify the effective scaling of the \fmo method. To this end, different clusters of substituted perylene bisimides (PBIs), where each monomer contains 142 atoms, were created, and the first 40 singlet excited states were calculated. Herefore, a basis of 2 locally excited states for each monomer and one charge-transfer state for each pair was employed. As shown in Fig. \ref{log_scaling}b, the asymptotic scaling of the FMO approach remained approximately the same, if the Davidson diagonalization and the electronic couplings are excluded. Since the PBI clusters consist of distinctly less monomers than the anthracene clusters, the Hamiltonian is also significantly smaller. Subsequently, the fit of the scaling, which is strongly determined by the timings of the bigger clusters, results in smaller factors for the PBI systems, and thus, the total run time only scales with a factor of $\mathcal{O}(N^{2.2})$. 
As the system size grows above 100 fragments (approx. 15000 atoms), the calculation of the Hamiltonian becomes the most time consuming factor of the whole calculation.
Subsequently, a further increase in the number of fragments would also result in a higher effective scaling factor as the Hamiltonian grows and the calculation of the couplings becomes the dominant aspect.

\subsection{Application to Pentacene} \label{ch:pentacene_application}
After demonstrating that \fmo can simulate excited states and absorption spectra with significantly reduced effort and, in return, at nearly the same accuracy as \tddftb, we wish to show the scope of possible applications by performing calculations on large pentacene clusters which serve as a model for bulk pentacene. This system has recently gained a lot of experimental and theoretical attention since it serves as a prototype for organic materials whose properties are determined by an interplay between local (Frenkel) and CT excitations.\cite{zimmerman_mechanism_2011,zirzlmeier_singlet_2015,beljonne_charge-transfer_2013}
The \fmo method was used to simulate the absorption spectrum of pentacene clusters of different sizes, which were generated from crystal structure data. As a first step, an adjustment of the long-range radius was required since the energies of the pentacene aggregates' charge-transfer states are overestimated with the default radius of 3.03 $a_0$. Subsequently, ionization potential and electron affinity (IP-EA) tuning\cite{stein_reliable_2009} were employed to estimate a reasonable value for the long-range radius by minimizing the following expression
\begin{equation}
    (\mathrm{IP}-\epsilon_{\mathrm{HOMO}})^2+(\epsilon_{\mathrm{LUMO}}-\mathrm{EA})^2.
\end{equation}
A scan of the radius yielded a minimum value of 6.875~$a_0$ as the optimal long-range radius which was used for all subsequent pentacene simulations. The ionization potential of 6.59~eV and the electron affinity of 1.35~eV were taken from experimental data of pentacene \cite{coropceanu_hole-_2002,crocker_electron_1993}. A value of 6.67~eV and 1.77~eV  was obtained using the optimally tuned radius for $\epsilon_{\mathrm{HOMO}}$ and $\epsilon_{\mathrm{LUMO}}$, respectively.
\begin{figure}[bth!]
    \centering
    \includegraphics[width=\linewidth]{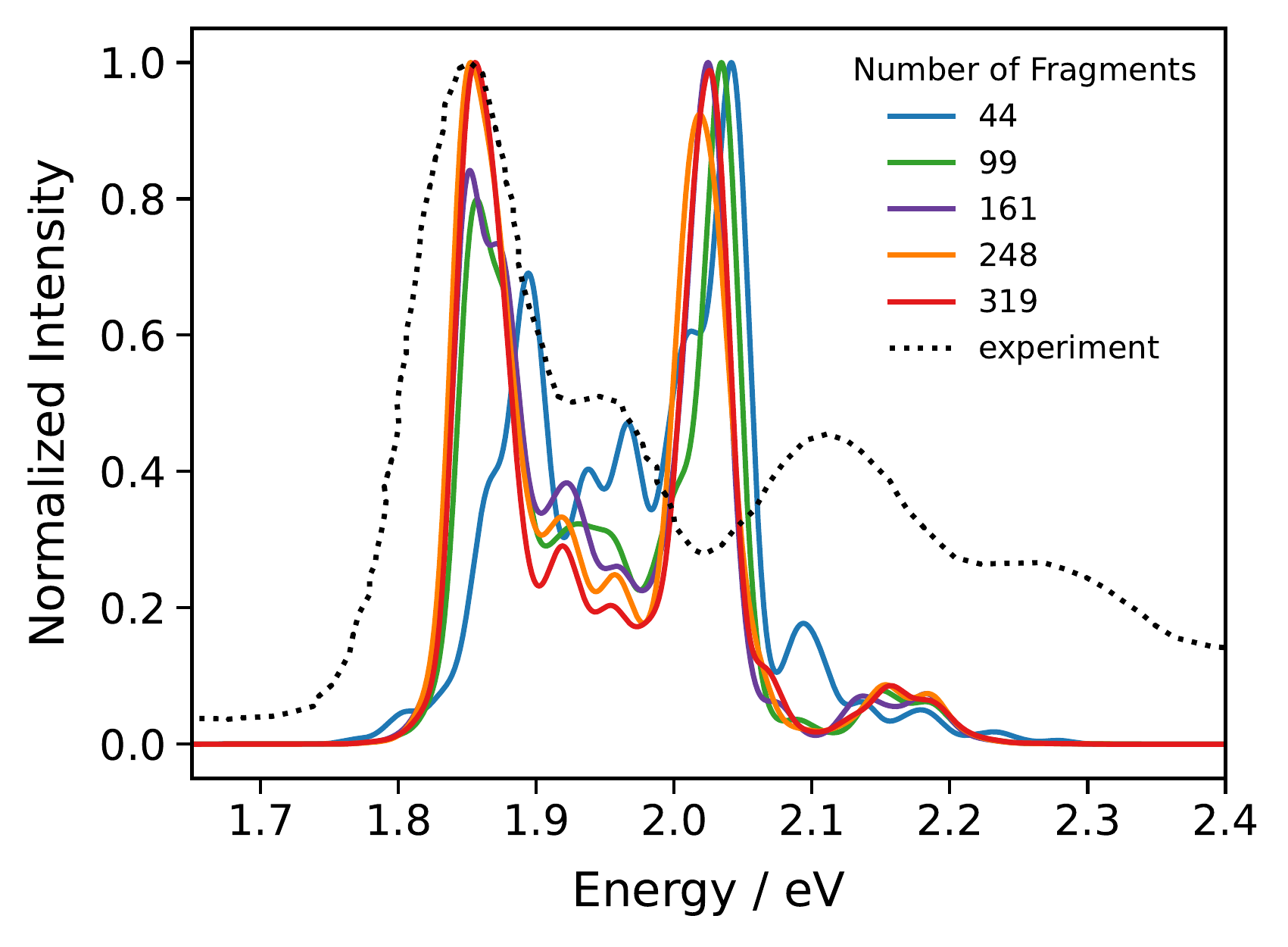}
    \caption{Comparison between the experimental spectrum of a pentacene film \cite{hammer_influence_2021} and the oscillator strengths of the \fmo calculation of pentacene clusters of various sizes. The energies of the clusters are red-shifted by 0.7 eV.}
    \label{pentacene_spectrum_vs_exp}
\end{figure}

For each of the pentacene clusters, which contain up to 319 fragments (11484 atoms), a calculation of the singlet excited state energies was performed. The basis states were constructed using three LE states for each monomer and one CT state for each pair. Depending on the system size, up to 3000 excited states have been calculated. The comparison between the experimental spectrum of a pentacene film\cite{hammer_influence_2021} and our results is shown in Fig, \ref{pentacene_spectrum_vs_exp}. The calculated oscillator strengths of the different pentacene clusters have been convolved with a Lorentzian line shape function of 10 meV width. In addition, the spectra were red-shifted by 0.7 eV to compensate for the error of the absorption energy. In the literature similar shifts are used even in the case of TD-DFT.\cite{green_excitonic_2021, hoche_excimer_2021, rohr_exciton_2018} 

Overall, the calculated oscillator strengths for the first two bands of the absorption spectrum are in good agreement with the experimental spectrum. As the size of the pentacene clusters increases, the intensity of the first band grows in strength, while the oscillator strength of the second band shows a slight decrease in magnitude. Comparing the third and fourth absorption bands of the spectra, the deviation in the energy and intensity are apparent. While the intensity of the third peak is overestimated by a factor of two, the magnitude of the forth band is severely underestimated by \fmo. However, these discrepancies may partially stem from the fact that the influence of the vibrational modes is fully neglected in the present theory, which play a significant role in the case of the pentacene clusters.\cite{benkyi_calculation_2019,craciunescu_accurate_2022, qian_herzbergteller_2020}

\begin{figure}[tbh!]
    \centering
    \includegraphics[width=\linewidth]{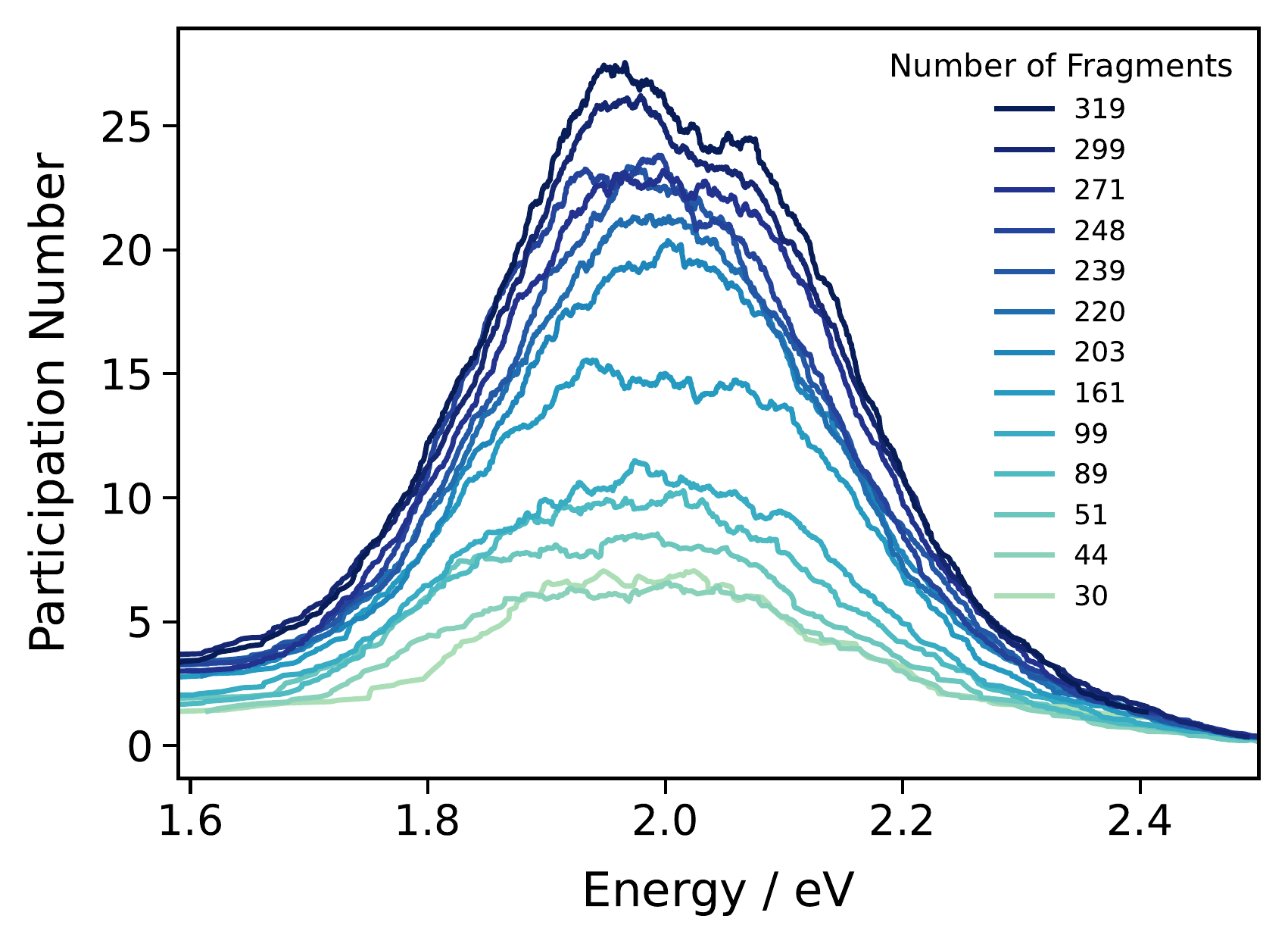}
    \caption{Average participation numbers of the excited states of the Wigner-sampled pentacene clusters.}
    \label{participation_numbers}
\end{figure}

Studying the excited states of large molecular systems leads to the question of how many monomers in a cluster contribute to an excited state. To this end, the participation numbers of the natural transition orbitals (NTOs)\cite{luzanov_application_1976,luzanov_interpretation_1980} of the excited states of the pentacene clusters were calculated according to the expression
\begin{equation}
    \mathrm{PR}_{\mathrm{NTO}}=\frac{\left(\sum_i \lambda_i\right)^2}{\sum_i \lambda_i^2},
\end{equation}
where $\lambda_i$ are singular values which were obtained by a singular value decomposition (SVD) of the transition density matrix between the electronic ground state and a singlet excited state \cite{plasser_new_2014}. Due to the fact that the pentacene monomers in the crystal structure all share the same internal coordinates, the locally excited states of the various monomers are nearly degenerate, which results in unrealistically high participation numbers for a perfectly arranged structure. In order to model the structural disorder that is present in a real system, we optimized the monomer geometry in the ground state using \dftb and sampled coordinates of the monomers from the ground-state canonical harmonic Wigner distribution.\cite{bonacic-koutecky_theoretical_2005}
Subsequently, the NTO participation numbers were calculated for the singlet excited states. In order to interpret the data, which showed strongly fluctuating participation numbers of the various states, the simple moving average was calculated by convolving the data with a rectangular window function. To guarantee a consistent approach, the width of the window was chosen to be 20\% of the complete number of data points of the respective system. The resulting participation numbers of the different pentacene clusters are shown in Fig.~\ref{participation_numbers}. 

As expected, as the size of the pentacene clusters increases, the number of monomers contributing to the excited states grows. However, a convergence of the participation numbers would also be expected after a specific system size is reached. 
In the case of the pentacene aggregates, the increase in the average participation number seems to be slowing down gradually.
\begin{figure}[tb!]
    \centering
    \includegraphics[scale=0.074]{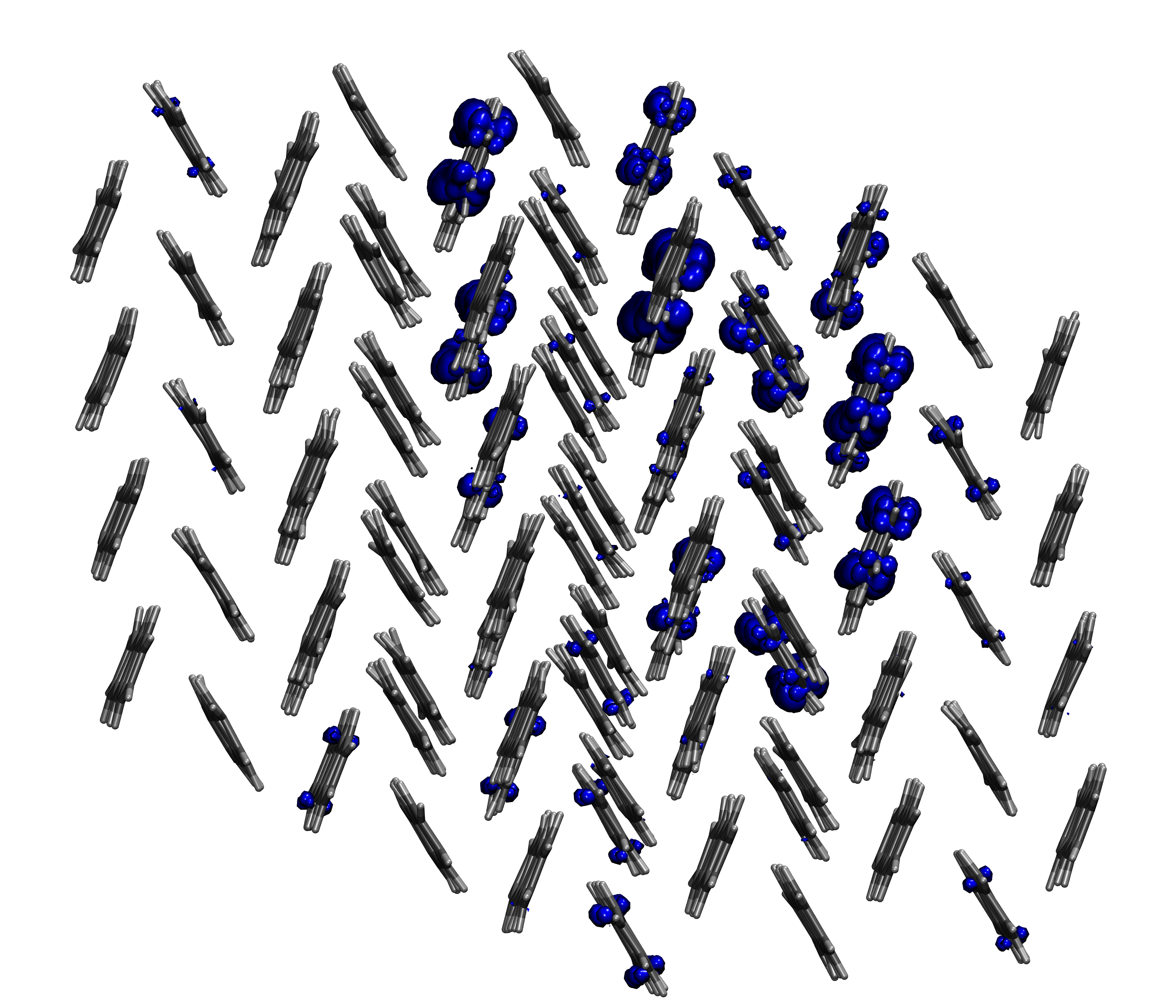}
    \includegraphics[scale=0.074]{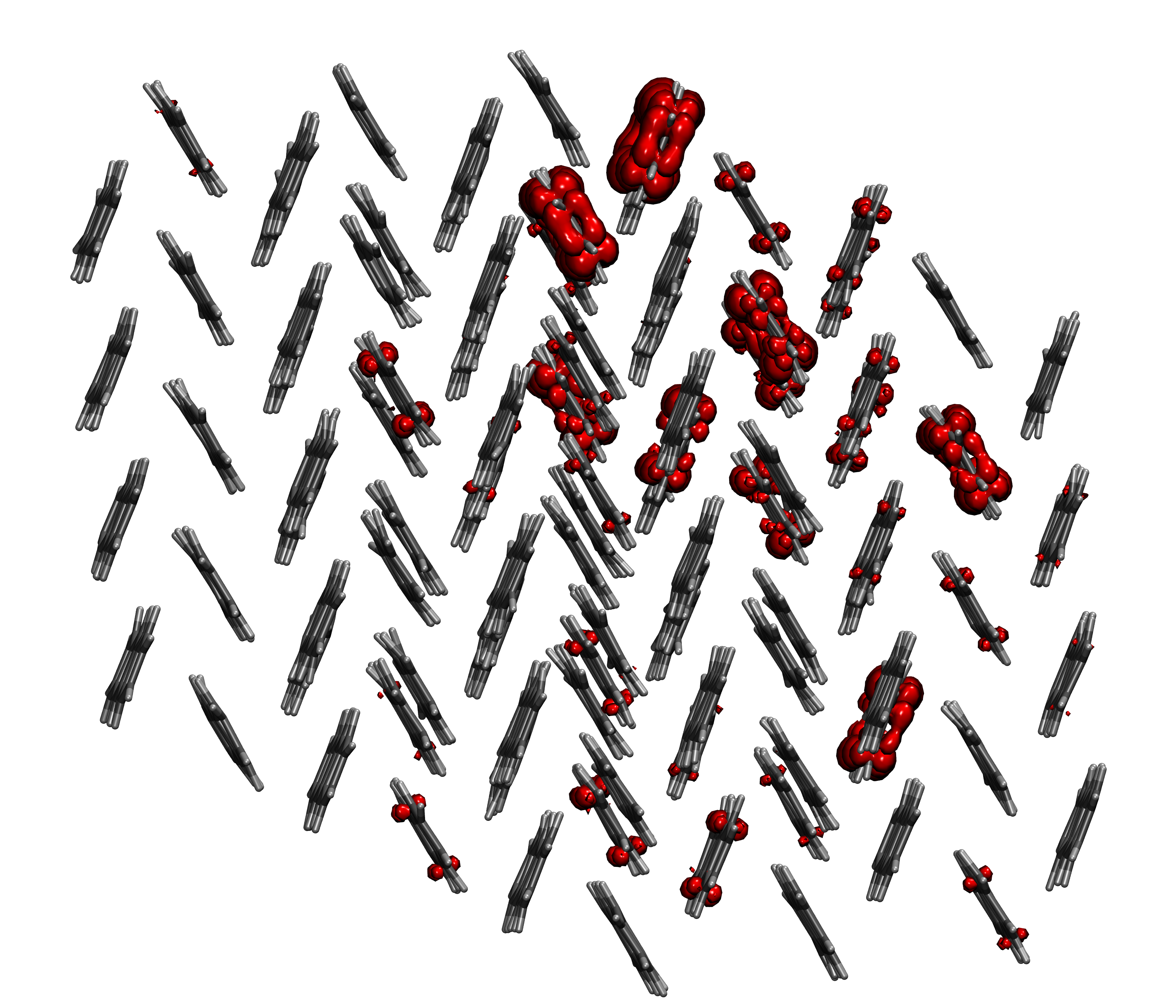}
    \caption{Particle and hole density (hole: blue, particle: red) of an excited state of the Wigner sampled pentacene cluster.
    The state represents a high participation number of the system. The isovalue of the plot is 0.0001.
    }
    \label{particle_hole_density}
\end{figure}

In order to visualize electron and hole delocalization the particle and hole density of a chosen pentacene system, containing 99 fragments (3564 atoms), was computed. The densities, which are shown in Fig. \ref{particle_hole_density}, were selected due to their relative large participation number of 22.7. Both local excitations and charge-transfer states are discernible.

Thus, we have proven the applicability of our method to large molecular systems. The \fmo method is sufficient to estimate the excited state properties of molecular aggregates and can be used to analyze the composition of the excited states.

\section{Conclusions and outlook}\label{sec:conclusions}
In this work, we have developed a new method to calculate the excited states in large molecular assemblies, consisting of hundreds of molecules. 
To this end, we based our approach on the fragment orbital density-functional tight-binding method\cite{nishimoto_density-functional_2014,vuong_fragment_2019} (FMO-DFTB) and employed an excitonic Hamiltonian constructed from locally excited and charge-transfer configuration state functions.
Our method has been implemented in the software package DIALECT, publicly available on Github\cite{noauthor_dialect_2022}.

We have proven that the accuracy of the proposed theory is sufficient to reproduce the excited state \tddftb potential energy curves of $\pi$--stacked molecular pyrene dimers. The mean absolute errors are below 80 meV at all physically relevant intermolecular separations for the pyrene systems, showing that for distances of 2.75 \AA \text{ } and above, our method produces satisfactory results. The applicability of the present theory was confirmed by the calculation of the excited state energies and oscillator strengths of a large anthracene cluster. The spectrum of the \fmo approach was in good agreement with the \tddftb reference. 

The comparison of the wall times of the \tddftb and \fmo calculation showed excellent speedup factors for different anthracene cluster sizes of up to $1.5\times 10^4$. In addition, the computational effective scaling of the proposed theory was evaluated for molecular aggregates involving differently sized monomers. The results showed that in the case of small monomers, the scaling is highly dependent on the calculation of the matrix elements of the Hamiltonian. However, if larger monomers like the substituted perylene bisimide are to be considered, the calculation of the matrix elements of the Hamiltonian only becomes the dominant factor of the effective scaling after the systems grows to a size of over 15000 atoms. The effective scaling of the anthracene and PBI systems was determined as $\mathcal{O}(N^{3.4})$ and $\mathcal{O}(N^{2.2})$, respectively.

At last, the application of our method to pentacene crystals confirmed its validity to large molecular systems. The calculated spectra of the pentacene clusters showed reasonable agreement to the experimental reference spectrum. We were able to analyze the number of monomers that contribute to the excited states of the system by calculating the NTO participation numbers, which were verified by the particle and hole density of a chosen pentacene cluster.

In the future, we plan to extend our approach to quantum-classical dynamics simulations of large molecular systems, like organic semiconductors or optoelectronic materials. To this end, we will implement the analytical gradients of the excited states to enable the investigation of exciton dynamics and charge transport simulations. Concerning the efficiency, graph theory methods will be used to accelerate the diagonalization of the Hamiltonian by partitioning the matrices involved in the Davidson algorithm into separate blocks, and thus reduce the dimensionality of the problem. In addition, we intend to apply the parametrization of the DFTB framework to a wider range of elements to facilitate the investigation of different molecular systems.

\section*{Conflicts of interest}
\vspace*{-2ex}
There are no conflicts to declare.
\section*{Acknowledgments}
\vspace*{-2ex}
We gratefully acknowledge financial support by the Deutsche Forschungsgemeinschaft via the grants MI1236/6-1 and MI1236/7-1.
\section*{Data availability}
\vspace*{-2ex}
The data that support the findings of this study are available on Github under \url{https://github.com/mitric-lab/Data_for_FMO-LC-TDDFTB}.
\section*{References}
\vspace*{-2ex}

\bibliography{references.bib}

\end{document}